\documentclass[twocolumn]{aastex63}

\def\a{\alpha}
\def\b{\beta}
\def\g{\gamma}
\def\y{\gamma}
\def\d{\delta}
\def\D{\Delta}
\def\e{\epsilon}
\def\i{\iota}
\def\k{\kappa}
\def\l{\lambda}
\def\u{\mu}
\def\v{\nu}
\def\s{\sigma}
\def\t{\tau}

\def\x{\xi}

\def\n{\nabla}
\def\w{\omega}
\def\z{\zeta}

\def\p{\phi}

\def\L{\Lambda}

\def\O{\Omega}

\def\apgt{\ {\raise-.5ex\hbox{$\buildrel>\over\sim$}}\ } 
\def\aplt{\ {\raise-.5ex\hbox{$\buildrel<\over\sim$}}\ }

\shorttitle{\textsc{55-min Period Eclipsing AM CVn}}
\shortauthors{Rodriguez and Galiullin et al.}
\graphicspath{{./}{figures/}}

\usepackage{float}
\usepackage{appendix}
\usepackage{xcolor}
\usepackage{amsmath}
\usepackage{ulem}

\begin{document}

\title{SRGeJ045359.9+622444: A 55-min Period Eclipsing AM CVn Discovered from a Joint SRG/eROSITA + ZTF Search}

\correspondingauthor{Antonio C. Rodriguez}
\email{acrodrig@caltech.edu}

\author[0000-0003-4189-9668]{Antonio C. Rodriguez}
\affiliation{Department of Astronomy, California Institute of Technology, 1200 E. California Blvd, Pasadena, CA, 91125, USA}

\author[0000-0001-5778-2355]{Ilkham Galiullin}
\affiliation{Kazan Federal University, Kremlevskaya Str.18, 420008, Kazan, Russia}

\author{Marat Gilfanov}
\affiliation{Space Research Institute, Russian Academy of Sciences, Profsoyuznaya 84/32, 117997 Moscow, Russia}
\affiliation{Max Planck Institute for Astrophysics, Karl-Schwarzschild-Str 1, Garching b. Muenchen D-85741, Germany}

\author{Shrinivas R. Kulkarni}
\affiliation{Department of Astronomy, California Institute of Technology, 1200 E. California Blvd, Pasadena, CA, 91125, USA}

\author{Irek Khamitov}
\affiliation{Kazan Federal University, Kremlevskaya Str.18, 420008, Kazan, Russia}
\affiliation{Academy of Sciences of Tatarstan Rep., Baumana Str. 20, Kazan 420111, Russia}

\author{Ilfan Bikmaev}
\affiliation{Kazan Federal University, Kremlevskaya Str.18, 420008, Kazan, Russia}
\affiliation{Academy of Sciences of Tatarstan Rep., Baumana Str. 20, Kazan 420111, Russia}

\author{Jan van Roestel}
\affiliation{Anton Pannekoek Institute for Astronomy, University of Amsterdam, 1090 GE Amsterdam, The Netherlands}

\author[0000-0003-2252-430X]{Lev Yungelson}
\affiliation{Institute of Astronomy, Russian Academy of Sciences, 48 Pyatnitskaya str., Moscow 109017, Russia}

\author{Kareem El-Badry}
\affiliation{Center for Astrophysics $|$ Harvard \& Smithsonian, 60 Garden Street, Cambridge, MA 02138, USA}
\affiliation{Department of Astronomy, California Institute of Technology, 1200 E. California Blvd, Pasadena, CA, 91125, USA}

\author{Rashid Sunayev}
\affiliation{Space Research Institute, Russian Academy of Sciences, Profsoyuznaya 84/32, 117997 Moscow, Russia}
\affiliation{Max Planck Institute for Astrophysics, Karl-Schwarzschild-Str 1, Garching b. Muenchen D-85741, Germany}

\author{Thomas A. Prince}
\affiliation{Division of Physics, Mathematics, and Astronomy, California Institute of Technology, Pasadena, CA 91125, USA}

\author{Mikhail Buntov}
\affiliation{Space Research Institute, Russian Academy of Sciences, Profsoyuznaya 84/32, 117997 Moscow, Russia}

\author{Ilaria Caiazzo}
\affiliation{Department of Astronomy, California Institute of Technology, 1200 E. California Blvd, Pasadena, CA, 91125, USA}

\author{Andrew Drake}
\affiliation{Department of Astronomy, California Institute of Technology, 1200 E. California Blvd, Pasadena, CA, 91125, USA}

\author{Mark Gorbachev}
\affiliation{Kazan Federal University, Kremlevskaya Str.18, 420008, Kazan, Russia}

\author{Matthew J. Graham}
\affiliation{Department of Astronomy, California Institute of Technology, 1200 E. California Blvd, Pasadena, CA, 91125, USA}

\author{Rustam Gumerov}
\affiliation{Kazan Federal University, Kremlevskaya Str.18, 420008, Kazan, Russia}
\affiliation{Academy of Sciences of Tatarstan Rep., Baumana Str. 20, Kazan 420111, Russia}

\author{Eldar Irtuganov}
\affiliation{Kazan Federal University, Kremlevskaya Str.18, 420008, Kazan, Russia}

\author{Russ R. Laher}
\affiliation{IPAC, California Institute of Technology, 1200 E. California
             Blvd, Pasadena, CA 91125, USA}

\author[0000-0002-8532-9395]{Frank J. Masci}
\affiliation{IPAC, California Institute of Technology, 1200 E. California
             Blvd, Pasadena, CA 91125, USA}

\author{Pavel Medvedev}
\affiliation{Space Research Institute, Russian Academy of Sciences, Profsoyuznaya 84/32, 117997 Moscow, Russia}

\author{Josiah Purdum}
\affiliation{Department of Astronomy, California Institute of Technology, 1200 E. California Blvd, Pasadena, CA, 91125, USA}

\author{Nail Sakhibullin}
\affiliation{Kazan Federal University, Kremlevskaya Str.18, 420008, Kazan, Russia}
\affiliation{Academy of Sciences of Tatarstan Rep., Baumana Str. 20, Kazan 420111, Russia}

\author{Alexander Sklyanov}
\affiliation{Kazan Federal University, Kremlevskaya Str.18, 420008, Kazan, Russia}

\author{Roger Smith}
\affiliation{Department of Astronomy, California Institute of Technology, 1200 E. California Blvd, Pasadena, CA, 91125, USA}

\author[0000-0003-4373-7777]{Paula Szkody}
\affiliation{Department of Astronomy, University of Washington, 3910 15th Avenue NE, Seattle, WA 98195, USA}

\author{Zachary P. Vanderbosch}
\affiliation{Department of Astronomy, California Institute of Technology, 1200 E. California Blvd, Pasadena, CA, 91125, USA}

\begin{abstract}

AM CVn systems are ultra-compact binaries where a white dwarf accretes from a helium-rich degenerate or semi-degenerate donor. Some AM CVn systems will be among the loudest sources of gravitational waves for the upcoming Laser Interferometer Space Antenna (LISA), yet the formation channel of AM CVns remains uncertain. We report the study and characterisation of a new eclipsing AM CVn, SRGeJ045359.9+622444 (hereafter SRGeJ0453), discovered from a joint SRG/eROSITA and ZTF program to identify cataclysmic variables (CVs). We obtained optical photometry to confirm the eclipse of SRGeJ0453 and determine the orbital period to be $P_\textrm{orb} = 55.0802 \pm 0.0003$ min. We constrain the binary parameters by modeling the high-speed photometry and radial velocity curves and find $M_\textrm{donor} = 0.044 \pm0.024 M_{\odot}$ and $R_\textrm{donor}=0.078 \pm 0.012 R_{\odot}$. The X-ray spectrum is approximated by a power-law model with an unusually flat photon index of $\Gamma\sim 1$ previously seen in magnetic CVs with SRG/eROSITA, but verifying the magnetic nature of SRGeJ0453 requires further investigation. Optical spectroscopy suggests that the donor star of SRGeJ0453 could have initially been a He star or a He white dwarf. SRGeJ0453 is the ninth eclipsing AM CVn system published to date, and its lack of optical outbursts have made it elusive in previous surveys. The discovery of SRGeJ0453 using joint X-ray and optical surveys highlights the potential for discovering similar systems in the near future. 
\end{abstract}

\section{Introduction}

AM Canum Venaticorum stars (AM CVn) are ultra-compact binaries, where a white dwarf (WD) accretes material from a helium-dominated, Roche lobe-filling donor. The orbital period of these systems lies in the 5.4--67.8 minutes range \citep[for recent reviews, see][]{2010solheim, 2018ramsay}. These ultra-compact binaries are potential laboratories to study accretion processes under extreme conditions. Due to the short orbital periods of AM CVn systems, their evolution is governed by the angular momentum loss (AML) via gravitational wave (GW) radiation \citep{1967AcA....17..287P}. \citet{1996ConPh..37..457H} recognized that AM CVn stars may be detected by space-based GW antennas. Indeed, many AM CVns are expected to be detected by LISA \citep[e.g.][]{2023LRR....26....2A, 2023lisa_kupfer}.  

AM CVns, like other cataclysmic variables (henceforth, CVs), originate from initially more widely separated binaries via common envelope evolution 
\citep{1976IAUS...73...75P}. Currently, three evolutionary channels are suggested for their
formation. In the {\it Helium WD donor} (He WD) channel \citep{1967AcA....17..287P}, two common envelope (CE) episodes leave behind a detached system composed by a more massive carbon-oxygen WD and a less massive helium WD. This system evolves to shorter orbital periods due to GW radiation. Once it reaches $P_\textrm{orb}\approx 5$ minutes, the WD donor fills its Roche lobe, and mass transfer begins. Since degenerate donors expand as they lose mass, the system evolves to longer orbital periods. 

Two CE episodes in the {\it Helium star} (He star) channel \citep{1987ApJ...313..727I} leave behind a  carbon-oxygen WD and a low mass He-star, which may be non-degenerate or weakly degenerate \citep{2023sarkarHEstar}. GW radiation brings the helium star to Roche lobe overflow. Mass transfer then results in the decrease of the orbital period to a minimum of about 10\,minutes, after which $P_\textrm{orb}$ starts to increase. Conventionally, binaries that evolved past $P_\textrm{orb}$ minimum are considered AM~CVn stars.

In the {\it evolved CV} channel, a single CE episode leaves behind a carbon-oxygen WD with a main-sequence companion. In this model, initially suggested by \citet{1985SvAL...11...52T}, the donor fills its Roche lobe when hydrogen in its core is almost exhausted ($X_c \aplt 0.1$). Such an inhomogeneous star becomes completely convective only when its mass decreases to several 0.01 $M_\odot$ and, hence, AML via magnetic braking may be at work. Later,  evolution is governed by GW emission. Such a system does not bounce at $P_\textrm{orb}\approx80$\,minutes like ``ordinary'' CVs, but may evolve to $P_\textrm{orb}\approx 30$\,minutes, if the \citet{1981A&A...100L...7V} empirical AML mechanism is assumed \citep[see ][]{2003MNRAS.340.1214P}. In the double-dynamo governed evolution model \citep{2023sarkarevolvedcv}, the driving force of evolution is magnetic braking due to the field generated in the boundary layer at $P_{\rm orb} \geq$ 3\,hr. Evolution is also governed by GW emission and the magnetic field, owing to differential rotation between core and outer layer at $P_{\rm orb} \leq 2$\,hr. In this model, the system may reach $P_{\rm orb} \approx$10 minutes. 

AM CVns exhibit a wide range of observed phenomena, often analogous to longer periods CVs with Roche lobe-filling main-sequence donors. In the He WD channel, the system immediately after contact may be so close that no accretion disk can form, and instead, direct impact accretion takes place. The possibility of stable mass exchange depends on the efficiency of tidal synchronization \citep{1984ApJ...277..355W,2001nelemans, 2004marsh,2011ApJ...737...89D}. Furthermore, at the high mass transfer rates that occur during direct impact accretion, the matter is ionized and no thermal instabilities can occur. Thus, systems with the lowest orbital periods do not show optical outbursts and can only be discovered through their periodic optical or X-ray emission \citep[e.g. HM Cancri;][]{2002hm_cnc}.

Systems with orbital periods in the range $P_\textrm{orb}\approx 20-50$ minutes transfer mass through an accretion disk and are subject to thermal instabilities \citep{2015cannizzo_outburst}. This leads to the vast majority of AM CVn systems being discovered through their optical outbursts and having periods in this range \citep[e.g.][]{2013levitan, 2021vanroestel}.  

Long-period AM CVn systems ($P_{\mathrm orb} \apgt $ 50 minutes) have such low mass transfer rates that outbursts are extremely rare. Furthermore, their low mass transfer rates make them optically dim and almost impossible to distinguish from WDs in a Hertzsrpung-Russell diagram.   These systems have only been identified in the past using two methods: 1) large-scale optical spectroscopic surveys with color cuts 
\citep{2013carter} and 2) eclipsing systems in all-sky photometric surveys \citep{2022vanroestel}. Each method has inefficiencies, as the former yielded only two new systems out of 2000 candidates; the latter approach is only sensitive to eclipsing systems and yielded five systems after performing a computationally expensive Box Least Squares \citep{kovacs_bls} search of 200,000 objects. Theoretical models suggest that the Galaxy may harbor up to $\simeq 3.4\times 10^7$ AM~CVn stars \citep[e.g.,][]{2004MNRAS.349..181N}, most of which should have long periods. Results of the above-mentioned  surveys may suggest that the models overpredict the number of Galactic AM CVns and they are really rare objects or that our current methods do not allow to detect majority of them. 

All-sky surveys provide unique possibilities to probe the Galactic population of compact object binaries. We are in the age of precise all-sky astrometry with Gaia and thousands of epochs of Northern sky photometric coverage with the Zwicky Transient Facility (ZTF) \citep{2016gaia, bellm2019}. While these surveys are useful on their own, they contain billions of targets. Such a large dataset is intractable without initial pre-selection, such as X-ray counterparts. CVs and AM CVns among them were discovered in the past through the all-sky Roentgensatellit X-ray survey \citep[ROSAT;][]{1982rosat1, 1999rosat2, 1997verbunt,  2002hm_cnc}. The ongoing eROSITA telescope aboard the Spektr-RG mission \textrm{\citep[SRG;][]{2021sunyaev, 2021erosita}} goes $\sim$15 times deeper than ROSAT, with improved localization of X-ray sources, and has already led to the discovery of new CVs \citep[e.g.][]{2022schwope, 2022AstL...48..530B, 2023rodriguez}. Some CVs were discovered using the Mikhail Pavlinsky ART-XC telescope \citep{2021A&A...650A..42P} on board of SRG observatory \citep[e.g.][]{2022A&A...661A..39Z}.

SRGeJ0453 is one of the sources discovered from a joint program to search for Galactic CVs from a cross-match of SRG/eROSITA and ZTF data in a $\rm 1200\ deg^2$ patch of the sky. The parameters and procedures of this CV search program will be described in full in  future publications. Its first results of the discovery and optical identification of several new Galactic CVs will be presented in upcoming work (Galiullin et al. in prep).  At the time of writing, no AM CVn system has been published using SRG/eROSITA data.

In this paper, we present the study and characterisation of the new long-period, eclipsing AM CVn, SRGeJ0453. In Section \ref{sec:data}, we present X-ray and optical observations with data reduction. In Section \ref{sec:discovery}, we outline the discovery and the optical and X-ray properties of SRGeJ0453. In Section \ref{sec:discussion}, we discuss our results, the binary parameters, the possible evolutionary channel of the donor, and compare it to the known population of AM CVns in the Hertzsprung-Russell diagram. We summarize our results in Section \ref{sec:summary}.

\section{Observations and data reduction}
\label{sec:data}

\begin{figure*}[t]
\begin{center}
\includegraphics[width=\textwidth]{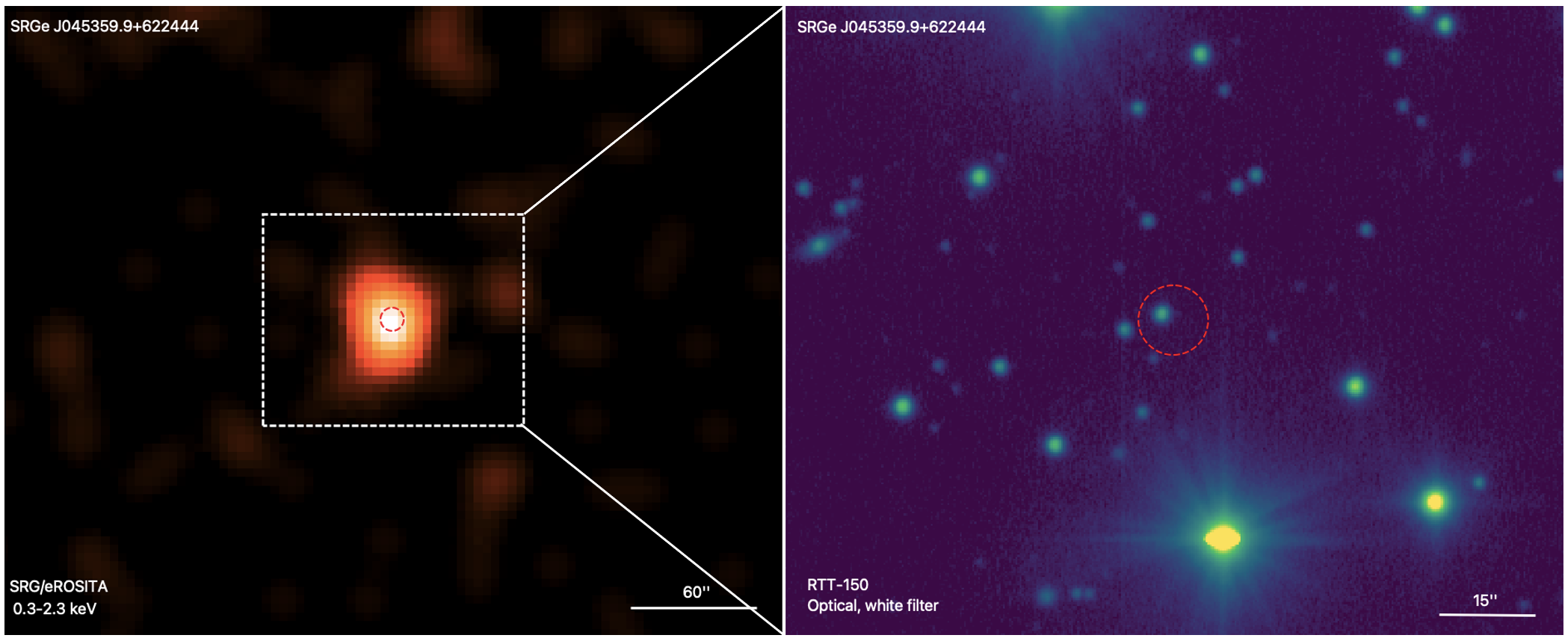}
\caption[] { {\it Left:} False-colour X-ray image of SRGeJ0453 in the 0.3--2.3 keV energy band from combined data of four all-sky surveys of SRG/eROSITA. The image was smoothed with a $15\arcsec$ Gaussian kernel.  {\it The white box} shows the field of view of the optical image on the right. {\it Right:} Optical image around SRGeJ0453 based on RTT-150 data with a white filter.  {\it The red circle} with a radius of  5.6$\arcsec$  (98\% localization error, R98) is centred at the X-ray position of SRGeJ0453.}
\label{fig:images}
\end{center}
\end{figure*}

\subsection{SRG/eROSITA}
On July 13, 2019, the Spektr--Roentgen--Gamma orbital observatory was successfully launched from the Baikonur Cosmodrome \citep{2021sunyaev,2021erosita}. SRG observatory carries two X-ray telescopes with grazing incidence optics: the ART-XC telescope, named after M. N. Pavlinsky \citep{2021A&A...650A..42P}, its operating range is 4--30 keV, and the eROSITA telescope, operating in the 0.2--8 keV energy band \citep{2021erosita}. We use all data collected by SRG/eROSITA in the survey mode between December 2019 and February 2022, which comprised $\approx4.4$ all-sky surveys. The initial data processing and calibration of the eROSITA data were carried out at the Space Research Institute of the Russian Academy of Sciences (IKI RAS) using the software developed in the eROSITA X-ray source catalogue science working group of the RU consortium with the use of the calibration tasks, and calibration database of the eSASS package (eROSITA Science Analysis Software) developed at the Max Planck Institute of Extraterrestrial Physics (MPE), Garching, Germany. The data were preprocessed using the results of ground pre-flight calibrations and flight calibration observations performed in October-November 2019 and throughout 2020--2022. The data from all sky surveys were combined to increase the sensitivity of the final source catalog. Figure \ref{fig:images} shows the X-ray image of the SRGeJ0453 obtained from combined data from four all-sky surveys of the SRG/eROSITA and an optical image of the same sky region from RTT-150.

To extract the X-ray spectrum of the source, we used a circle with a radius of $40\arcsec$, and an annulus with inner and outer radii of $120\arcsec$ and $300\arcsec$, respectively, for the background region. The X-ray spectrum was analysed using XSPEC v.12 software \citep{1996ASPC..101...17A}. Due to the low number of source spectrum counts, we approximated it using the $C$-statistics \citep{1979ApJ...228..939C}. The spectrum was grouped to have at least three counts per spectral channels\footnote{See note for work in XSPEC: \url{https://heasarc.gsfc.nasa.gov/xanadu/xspec/manual/XSappendixStatistics.html}}.

\subsection{ZTF}
The Zwicky Transient Facility (ZTF) is a photometric survey that uses a wide 47 $\textrm{deg}^2$ field-of-view camera mounted on the Samuel Oschin 48-inch telescope at Palomar Observatory with $g$, $r$, and $i$ filters \citep{bellm2019, graham2019, dekanyztf, masci_ztf}. In its first year of operations, ZTF carried out a public nightly Galactic Plane Survey in $g$-band and $r$-band \citep{ztf_northernskysurvey_bellm, kupfer_ztf}. This survey was in addition to the Northern Sky Survey which operated on a 3 day cadence \citep{bellm2019}. Since entering Phase II, the public Northern Sky Survey is now at a 2-day cadence. The pixel size of the ZTF camera is 1$\arcsec$ and the median delivered image quality is 2.0$\arcsec$ at FWHM. 

We use ZTF forced photometry taken through February 1, 2023, processed by IPAC at Caltech\footnote{\url{https://irsa.ipac.caltech.edu/data/ZTF/docs/ztf_forced_photometry.pdf}}. Light curves have a photometric precision of 0.015--0.02$^m$ at  14$^m$ down to a precision of 0.1--0.2$^m$ for the faintest objects at 20--21$^m$.  While both the default photometry and forced photometry from ZTF use PSF-fit photometry, the forced photometry calculates the flux of the object on difference images by \textit{forcing the location of the PSF to remain fixed} according to the ZTF absolute astrometric reference. This allows one to obtain flux estimates below the detection threshold and therefore probe deeper than the standard photometry. 

\begin{figure}
    \centering
    \includegraphics[width=0.45\textwidth]{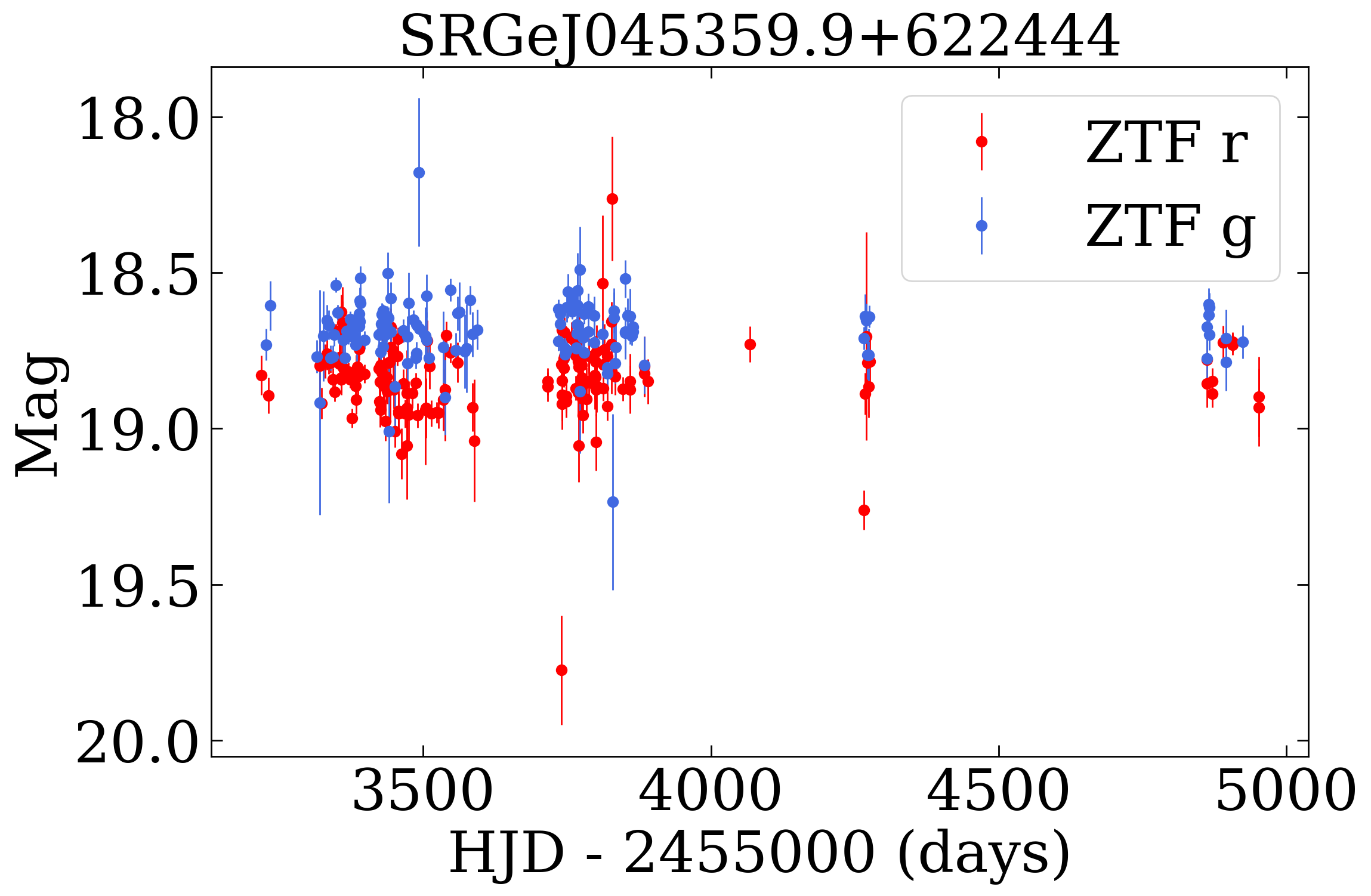}\\
    \includegraphics[width=0.43\textwidth]{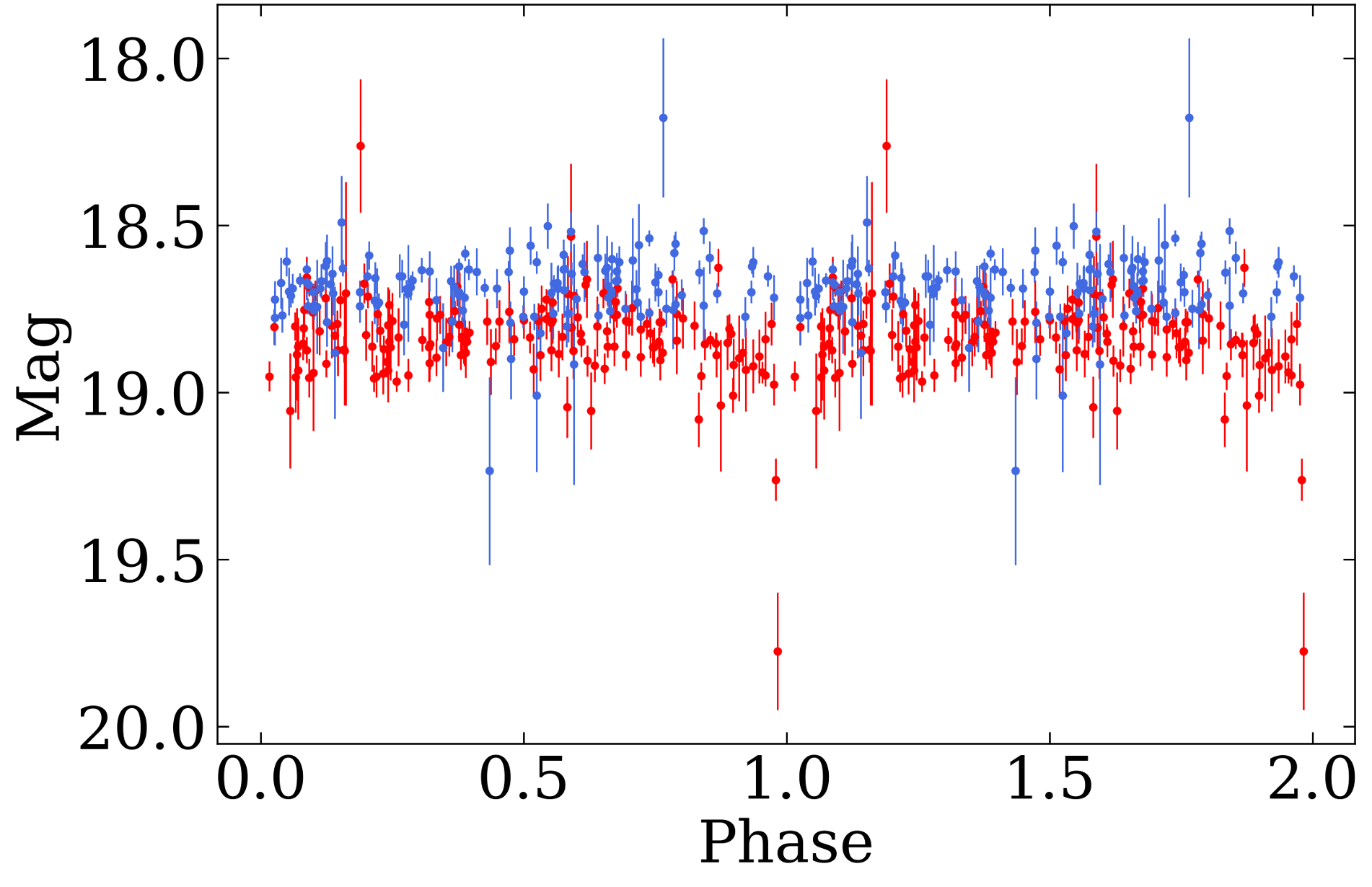}
    \caption{ZTF light curves of SRGeJ0453 on $g$, $r$ filters: long-term (top), and folded at the 55.08 minutes orbital period (bottom). No significant outbursts are seen over the $\approx$5 yr-long baseline.}
    \label{fig:ztf_lc}
\end{figure}

We present archival ZTF data (SNR $>$ 5; no upper limits shown) for SRGeJ0453 in Figure \ref{fig:ztf_lc}. The lack of optical points is due to the object consistently falling on the edge of a ZTF camera chip. This coincidence led to poor quality data taken after JD 2459000 and a subsequent deficiency in observing epochs. 

\subsection{RTT-150 photometry}
\label{sec:rtt150}

\begin{figure}[t]
\begin{center}
{
\includegraphics[width=0.47\textwidth,clip=true] {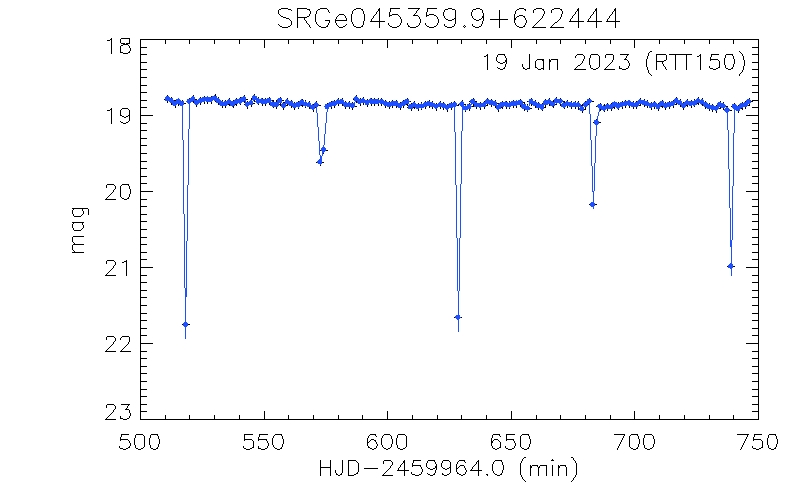}
\includegraphics[width=0.47\textwidth,clip=true]{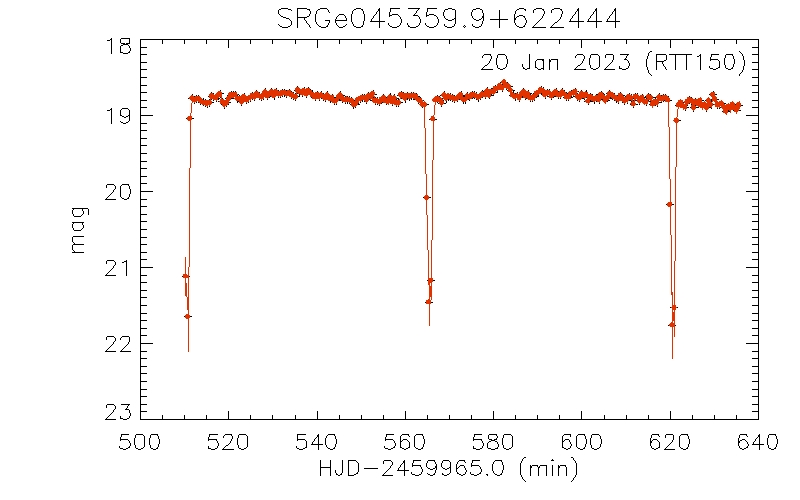}
}
\caption[] {Optical light curve of SRGeJ0453 observed by RTT-150, where axes are G-magnitude versus the time from the start of the observations in units of minutes (${\it top}$ -- the first; and ${\it bottom}$ -- the second nights of observations).
}
\label{fig:lc_rtt150}
\end{center}
\end{figure}

We observed SRGeJ0453 with the 1.5-meter Russian-Turkish Telescope (RTT-150) as part of the optical follow-up program of SRG/eROSITA sources. The observations were performed using the TFOSC instrument ANDOR CCD camera, model iKon-L 936 and BEX2-DD-9ZQ (a $2048 \times 2048$ pixels) chip. The resolution element is 0.326$\arcsec$ at $1\times 1$ binning. Two sets of observations were carried out on January 19 and 20, 2023. The weather was clear, and the average seeing was 1.5--2.0$\arcsec$. All images were pre-processed using ATROLIB/IDL packages and the standard method: bias subtracted and flat-field corrected. In addition, astrometric alignment was performed for all images and forced photometry.

A total of 160 images (with 60 seconds of exposure time and 28 seconds of readout time) were obtained on January 19 with a sub-frame size of 2048 x 600 pixels and binning 1 x 1. The total duration of observations was 245 minutes, and the time resolution was 88 seconds on the first night of observations. Five minima (eclipses) were detected with different depths of minima. Two possible periods were found (55 or 110 minutes) by using the first night's light curve. We found that eclipses were very narrow ($\approx$ 60 seconds); therefore, different depths of minima were produced due to insufficient time resolution during the first night observations. On the second night (January 20, 2023), 258 images with 20 seconds of exposure time and 9 seconds of readout time were obtained with the sub-frame size 1024 x 200 pixels (binning 2$\times$2). The total duration of observations was 135 minutes, and the time resolution was 29 seconds. The three minima (eclipses) were detected with the same depths of minima, confirming 55 and excluding the 110 minutes period. 

\begin{figure}
    \centering
    \includegraphics[width=0.47\textwidth]{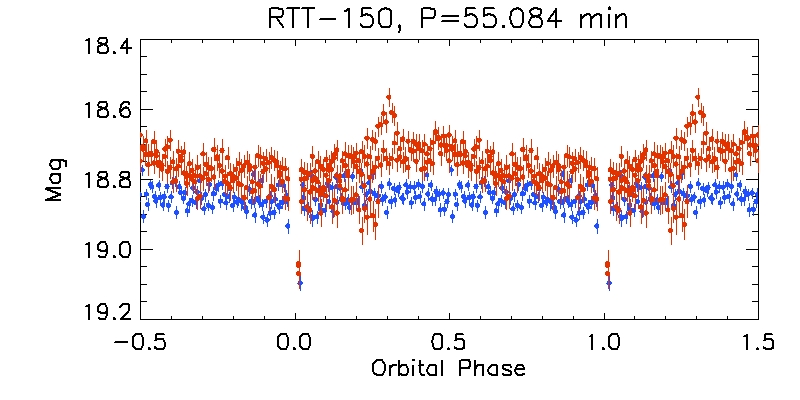}
    
    \caption{Phase-folded light curve of RTT-150 data. $\it{Blue~circles}$ corresponds to January 19, 2023 data, $\it{red~circles}$ -- 
    January 20, 2023 data. The light curve shows low amplitude ($\approx0.1-0.3^m$) flickering. Only the upper part of the light curve is shown out of eclipses.}
    \label{fig:RTT150_flc_all}
\end{figure}

Having eight minima within both nights' light curves, we searched for an accurate period by using a combined approach (for more details, see Appendix \ref{app:rtt150_period}). We found the period to be $55.084 \pm 0.015$ minutes. The optical light curves of  SRGeJ0453 obtained by RTT-150 are shown in Figures \ref{fig:lc_rtt150} (long term), and Figure \ref{fig:RTT150_flc_all} (folded at the orbital period).

\subsection{LRIS spectroscopy}

We obtained an identification spectrum of SRGeJ0453 on the Keck I telescope using the Low-Resolution Imaging Spectrometer \citep[LRIS;][]{lris} on January 15, 2023 (UT). We used the 600/4000 grism on the blue side with 2x2 binning (spatial, spectral), and the 600/7500 grating on the red side with 2x1 binning. We used a 1.0$\arcsec$ slit, and the seeing during the portion of the night was approximately 0.7$\arcsec$, leading to minimal slit losses. Phase-resolved spectroscopy over the entire orbit was acquired on March 25, 2023 (UT) using LRIS. In that case, the 600/4000 grism (2x2 binning) was used on the blue side, and the 400/8500 grating on the red side (2x1 binning). A 1.0$\arcsec$ slit was used, and the seeing during the portion of the night was approximately 0.9$\arcsec$. Wavelength coverage and resolution are shown in the timeline of all observations in Table \ref{tab:data}.

On the night of  March 25, 2023, observations were affected by telescope vignetting, resulting in a loss of throughput. In early January 2023, the Keck I bottom shutter drive sustained damage, forcing it to remain stationary at its park position of 24 degrees elevation. In this condition, all observations taken below an elevation of 38 degrees are affected by vignetting. That night, all 6 exposures were forced to be taken at an elevation below 30 degrees (corresponding to airmass $\approx 1.9$) as the object was quickly setting. 

All Keck I/LRIS data were reduced with \texttt{lpipe}, an IDL-based pipeline optimized for LRIS long slit spectroscopy and imaging \citep{2019perley_lpipe}. All data were flat fielded and sky-subtracted using standard techniques. Internal arc lamps were used for the wavelength calibration and a standard star for overall flux calibration. 

\subsection{CHIMERA high-speed photometry}

\begin{figure}
    \centering
    \includegraphics[width=0.45\textwidth]{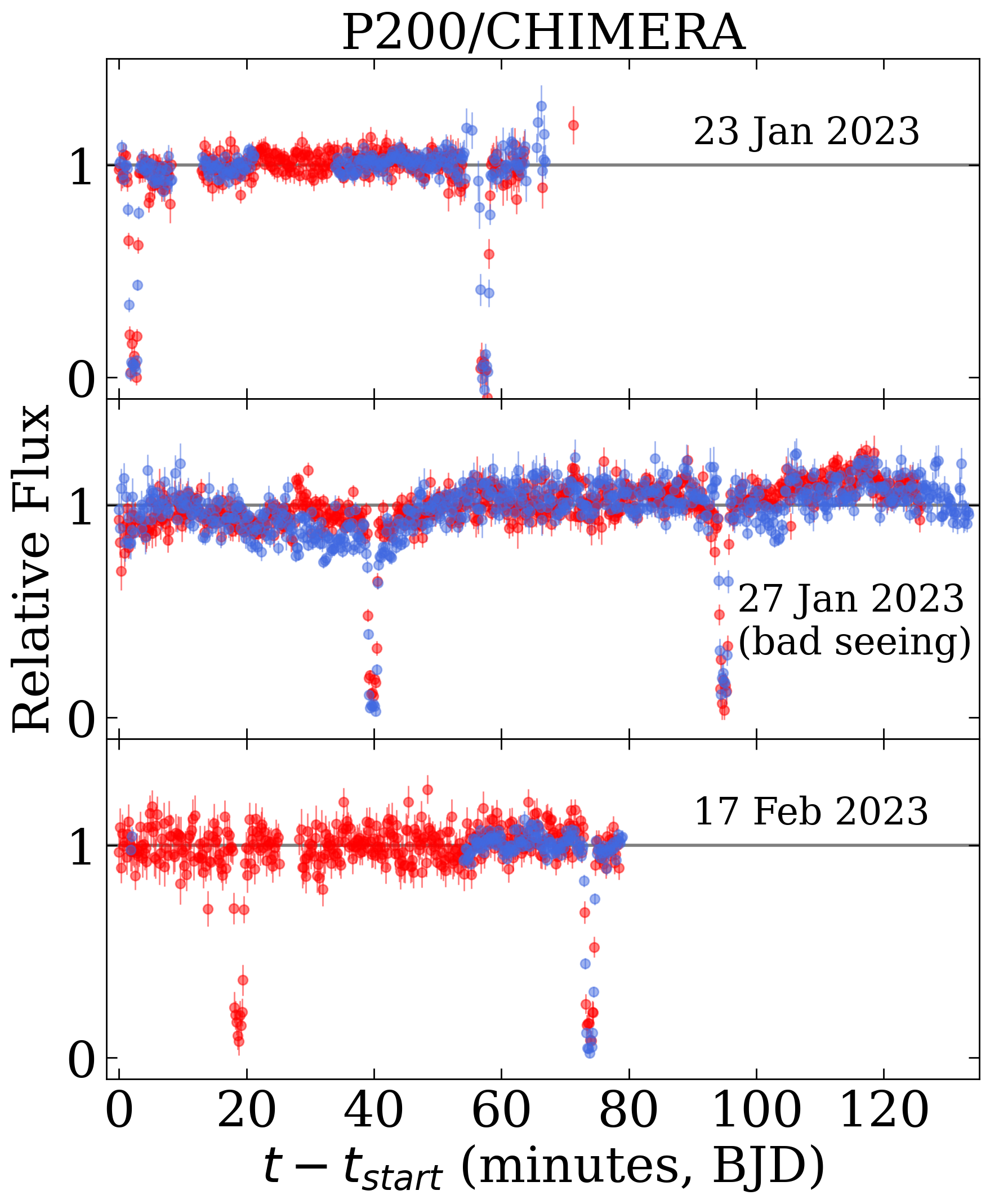}
    \includegraphics[width=0.5\textwidth]{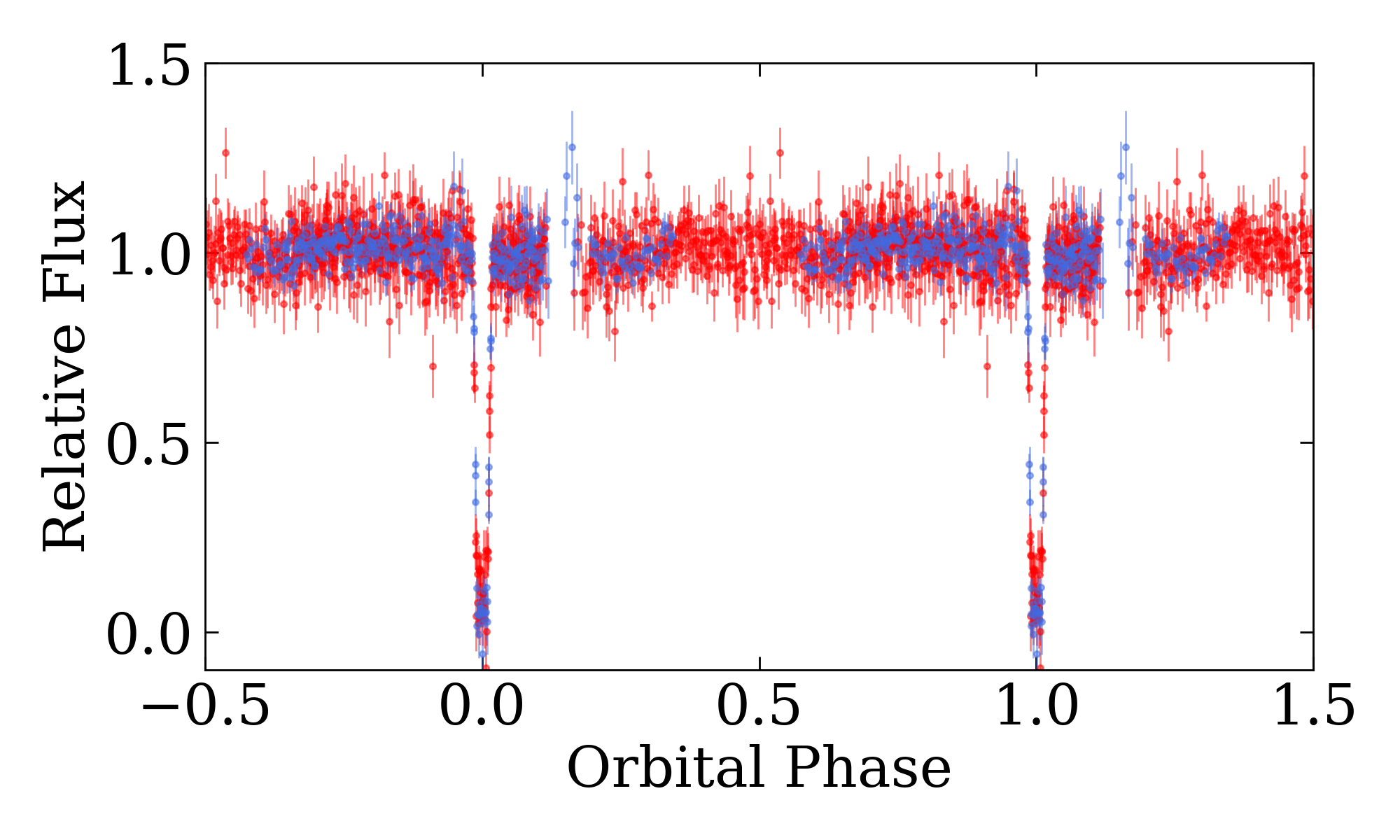}
    \caption{CHIMERA $r$ (red) and $g$ (blue) 10-sec cadence photometry reveal deep eclipses. \textit{Upper three panels}: The entire observation on each occasion. Gaps are due to large error bars in the data where cloud cover or highly variable seeing prevented a good extraction of the data. \textit{Bottom panel}: Data from January 23 and February 17, 2023 are folded over the 55.08 minutes orbital period.}
    \label{fig:CHIMERA}
\end{figure}

We acquired high-speed photometry in Sloan $r$ and $g$ bands using the Caltech HIgh-speed Multi-color camERA \citep[CHIMERA;][]{chimera} on three occasions. In all cases, CHIMERA $r$-and $g$-band data were both acquired at a 10 second cadence simultaneously over two orbital periods. We are clearly able to identify eclipses in all CHIMERA data sets, but it may be difficult to interpret out of eclipse variability due to abnormally poor seeing on every night: 3$\arcsec$ on January 23, 2023 (UT), 4$\arcsec$  on January 27, 2023 (UT), and 3$\arcsec$  on February 17, 2023 (UT). Data from the night of January 27, 2023 (UT) was most strongly affected by poor seeing, and the overall flux level surrounding the eclipse was affected. In all the light curve analysis, we only use data from January 23, 2023 (UT) and February 17, 2023 (UT). We show all light curves in Figure \ref{fig:CHIMERA} and summarize this information in Table \ref{tab:data}.

All CHIMERA data were bias-subtracted and flat-fielded using standard techniques in the \texttt{PyCHIMERA} pipeline\footnote{\url{https://github.com/caltech-chimera/PyChimera}}. The ULTRACAM pipeline was used to do aperture photometry \citep{2007ultracam}. A differential light curve was created by dividing the counts of the target by those of a standard star. We experimented with different standard stars to ensure we were not affected by standard star variability.  

\begin{deluxetable*}{ccccc}
\label{tab:data}
 \tablehead{
 \colhead{Data Type} & \colhead{Date (UT)} & \colhead{Instrument} & \colhead{Specifications} & \colhead{Finding}  }
 \tablecaption{Data Acquired for SRGeJ0453}
 \startdata\hline
Identification Spectrum & 15 Jan. 2023 &{\parbox{4cm}{\vspace{5pt}\centering Keck I/LRIS}}& {\parbox{4cm}{\centering \vspace{5pt}Blue: 3140--5640 \AA, \\$\Delta \lambda = $1.1 \AA, 1x900s exp.  \\ Red: 5530--8830 \AA,\\ $\Delta \lambda = $0.80 \AA, 1x900s exp.}}&{\parbox{4cm}{\vspace{5pt}\centering Helium emission lines over a blue continuum; Central absorption suggesting high inclination}}\\\hline
{\parbox{2cm}{\centering \vspace{5pt} Photometry, no filter}}
 & 19, 20 Jan. 2023 & {\parbox{4cm}{\vspace{5pt}\centering RTT-150/TFOSC }}&{\parbox{4cm}{\centering \vspace{5pt} 60s and 20s exp. for 4.1 and 2.2 hrs }} & {\parbox{4cm}{\vspace{5pt}\centering Photometry revealed deep (3 mag) eclipse and 55.084 minutes period}}\\\hline
{\parbox{2cm}{\centering \vspace{5pt}High-cadence $r$ and $g$ band Photometry}}
 & {\parbox{4cm}{\vspace{5pt}\centering 23,27 Jan. 2023 \\
 17 Feb. 2023}}  & {\parbox{4cm}{\vspace{5pt}\centering Hale Telescope/CHIMERA}}&{\parbox{4cm}{\centering \vspace{5pt}10s exp. for 2 hr }} & {\parbox{4cm}{\vspace{5pt}\centering High-cadence photometry at simultaneous orbital phases}}\\ \hline
Multi-phase spectra & 25 Mar. 2023 &{\parbox{4cm}{\vspace{5pt}\centering Keck I/LRIS}}& {\parbox{4cm}{\centering \vspace{5pt}Blue: 3240--5640 \AA, \\$\Delta \lambda = $2.0 \AA, 6x600s exp.  \\ Red: 5530--10,200 \AA,\\ $\Delta \lambda = $1.2 \AA, 6x600s exp.}}&{\parbox{4cm}{\vspace{5pt}\centering Doppler tomograms reveal bright spots and ``central spike" in He II 4686}}\\
\enddata

\end{deluxetable*}

\section{Discovery and Results}
\label{sec:discovery}

SRGeJ0453 is one of the objects identified as a CV candidate in a cross-match of a $\rm 1200\ deg^2$ patch of sky of SRG/eROSITA X-ray data with Gaia proper motion data and the optical ZTF database.  SRGeJ0453 was called to our attention by its proper motion statistically significantly detected by Gaia, high ratio of X-ray flux to optical flux, $\rm F_{X}/F_{opt}\approx0.12$, and placement in the Gaia color-magnitude diagram near the WD region. This is a pilot study, and more targets identified from this program will be presented in future work (Galiullin et al. in prep.).

\subsection{ Period Determination }
\label{sec:period}
We detected  eight deep eclipses in two observing nights with RTT-150, determining  the system's orbital period as $55.084 \pm 0.015$ minutes (see Section \ref{sec:rtt150}). The light curve shows low amplitude ($\approx(0.1-0.3)^m$) flickering, possibly caused by an accretion disk. During eclipses, the light curve shows deep dips ($\approx3^m$) separated by $\approx$ 55.08 minutes (see Figures \ref{fig:lc_rtt150} and  \ref{fig:RTT150_flc_all}).

We ran the Box Least Squares (BLS) algorithm \citep{kovacs_bls}  on the ZTF data to find the orbital period of SRGeJ0453. We used the exposure time as an oversampling factor when creating our frequency grid to search for possible periods and durations of the eclipse. We found the best-fit period of 55.1$\pm$0.5 minutes using the ZTF forced photometry data, where the uncertainty is determined by exposure time. The ZTF optical light curves of SRGeJ0453 folded with a best-fit period are shown in Figure \ref{fig:ztf_lc}.

To clarify and better constrain the period of  SRGeJ0453, we ran the BLS algorithm on all CHIMERA light curves, recovering a period of 55.08$\pm$0.08 minutes. Starting with the 55.08 minutes period determined from a single CHIMERA observation, we assume there is no significant period derivative and calculate precisely 652 orbits to transpire between the first (January 23, 2023) and last eclipses (February 17, 2023) measured with CHIMERA.  By matching the observed mid-eclipse time of all CHIMERA eclipses to the expected mid-eclipse time within 10 seconds (exposure time of individual CHIMERA exposure), we can obtain a much more precise period measurement. Using this method, we obtain a period estimate of $P_\textrm{orb} = 55.0802 \pm 0.0003$ minutes. We present all good quality CHIMERA data folded on this period in Figure \ref{fig:CHIMERA}. 

All periods estimated from different optical data agree with each other within their respective error bars. In all analysis, we adopted the orbital period based on CHIMERA and RTT-150 light curves $P_\textrm{orb}\approx$ 55.08 minutes, and ephemeris $t_0$(BJD)=  2459967.726794129(1).

\subsection{Optical Spectral Features}

\begin{figure*}
    \centering
    \includegraphics[width=\textwidth]{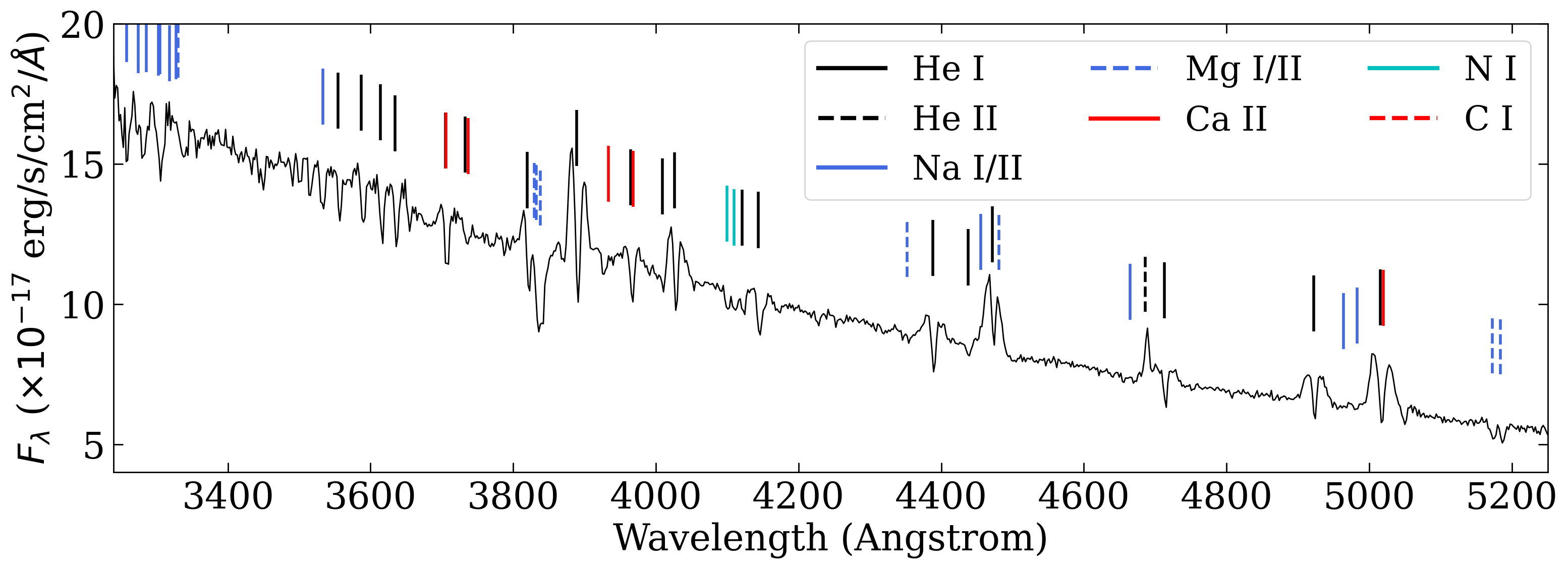}\\\includegraphics[width=\textwidth]{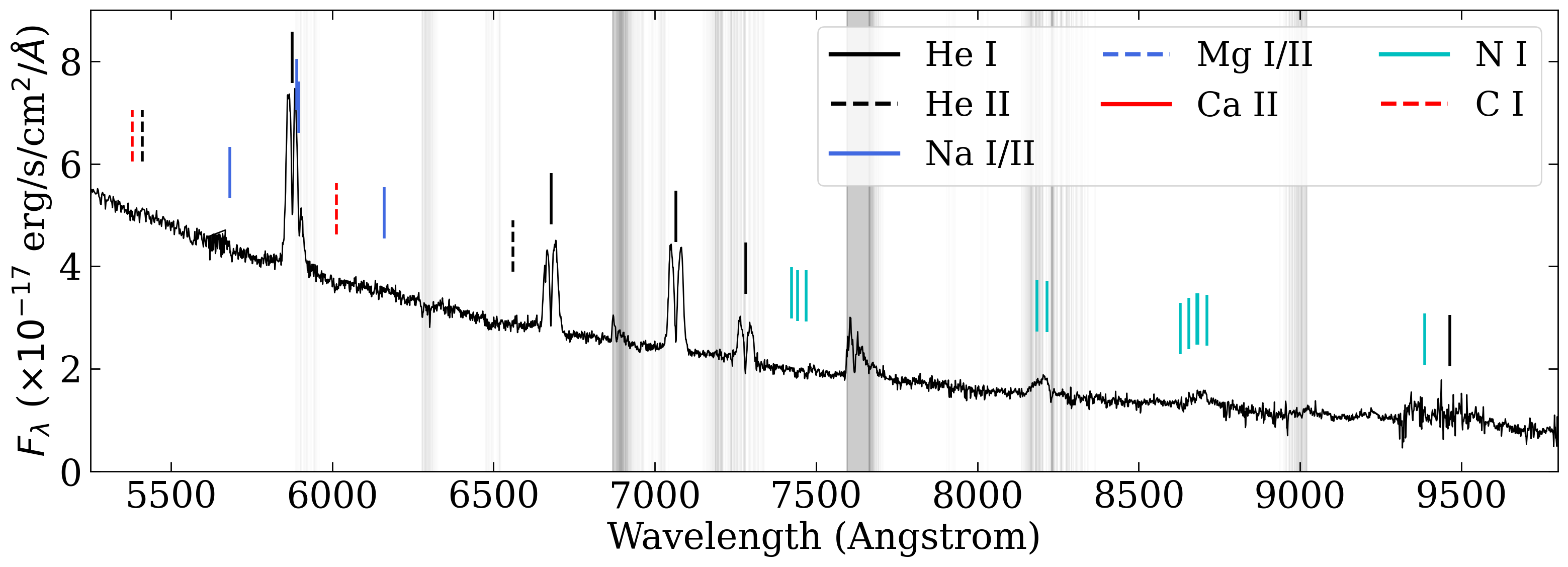}
    \caption{LRIS phase-averaged spectrum of SRGeJ0453. The characteristic features of an AM CVn are clear: helium emission lines imposed over a blue continuum. Metals such as Mg, Ca, N, and Na are present. Grey lines are locations where there are telluric features from the Keck Telluric Line List.}
    \label{fig:LRIS}
\end{figure*}

\begin{table}
	\centering
	\caption{Equivalent widths (EWs) of selected lines
	}
	\label{tab:EW_lines}
	\begin{tabular}{lc} % four columns, alignment for each
		\hline
		Line (\AA) &   {\parbox{2cm}{\vspace{5pt}\centering LRIS 25 Mar. EW (\AA)}} \\
		\hline
\textit{Emission features} &\\
He I\ 4387.9 & $-1.68\pm 0.02$\\
He I\ 4921.9 & $-3.0\pm 0.1$\\
He I\ 5876.5  & $-28.7\pm 0.3$\\
He I\ 6678.2 & $-24.5\pm 0.1$\\
He I\ 7065.2  & $-34.3\pm 0.2$\\
He I\ 7281.4  & $-11.6\pm 0.5$\\
He II\ 4685.7  & $-1.7\pm 0.1$\\\textit{Absorption features}  &\\
C I\ 6013.2 & $0.2\pm 0.2$\\
N I\ 7442.3  & $0.8\pm 0.2$\\
N I\ 7468.3 & $0.5\pm 0.1$\\
Na II\ 3304.96  & $1.35\pm 0.03$\\
Mg I\ 5183.6  & $1.3\pm 0.2$\\
Ca II\ 3933.7 & $0.6\pm 0.1$\\
\hline
	\end{tabular}
\end{table}

We present the optical spectrum of SRGeJ0453 in Figure \ref{fig:LRIS}. We show the average of the first five spectra taken on March 25, 2023 with LRIS, which cover an entire orbit (with approximately 10 percent gap due to read out time). The phase coverage of the sixth spectrum overlaps with that of the first, and was taken near the limiting elevation of 24 degrees, so it is omitted from all analysis. 

The archetypal features of an AM CVn system are clear: a blue continuum with He lines (in emission and/or absorption) and an absence of H lines. There is no Balmer jump (or inverse thereof, as commonly present in CVs) in the spectrum. All He I emission lines redward of 3800 \AA \;are clearly doubled, originating from the helium rich accretion disk around the WD. All He I emission lines show a narrow central absorption commonly seen in nearly edge-on (high inclination) AM CVns. This phenomenon has been explained by WD light passing through an accretion disk that is optically thick at central He I wavelengths, but optically thin in the wings. 

We see a single-peaked emission line of He II 4685.7 \AA, which we discuss in detail in Section \ref{sec:central_spike}. Table \ref{tab:EW_lines} shows equivalent widths (EWs) for prominent lines identified in the optical spectrum of SRGeJ0453, which we calculate from the averaged spectrum.

Metal lines are also present in the spectrum, which can tell us about the formation channel of this system. We present the most prominent elements below:
\begin{itemize}
    \item C: There is at best marginal ($1\sigma$) evidence for C~I 6013.2 \AA\; in absorption in the phase-averaged spectrum of SRGeJ0453. The dip in the spectrum at that position is around 5 percent below the continuum, which is only marginally above the noise level. Therefore, we do not claim this as a significant detection of carbon in SRGeJ0453. No other C I or C II lines are present. 
    \item N: We see N I in both emission (8629.2, 8655.8, 8680.28, 8683.4, 8711.7 \AA) and absorption (7442.3, 7468.3 \AA). The complex around 8184.9, 8188.0, and 8216.30 \AA\ are all near telluric features, making their clear identification difficult.
    \item O: No O I or O II lines are present.
    \item Other metals: We detect Na I/II lines, Mg I/II, and Ca II lines, all in absorption. 
\end{itemize}

With a confident detection of various lines of nitrogen, we are able to place constraints on the formation channel of SRGeJ0453. The lack of oxygen and the high nitrogen to carbon ratio favors either the He star or He WD formation channel, and makes the evolved CV channel unlikely. A detailed discussion and connection to formation channel is in Section \ref{sec:evolution}.

\subsection{Distance, Extinction, and $N_H$}
The only Gaia source associated with SRGeJ0453 within 5.6$\arcsec$ search radius (98\% localization error, R98) has an ID 477829370972112000 (Gaia EDR3) and celestial coordinates of RA=$04^{h} 54^{m} 00^{s}.1$ and DEC=$+62\degr 24^{'} 45^{''}.6$. The distance calculated using Gaia EDR3 parallaxes is $d = 239^{+11}_{-8}$ pc \citep{2021bailerjones}. We adopt an extinction value of $E(B-V) = 0.01\pm 0.004$ using the Bayestar19 dust map \citep{bayestar19}. We use the extinction law of \cite{ccm1989} and a value of $R_V = 3.1$ to calculate the flux correction from UV to IR wavelengths. We can now calculate the intervening hydrogen column, $N_H = (6.8 \pm 2.8) \times 10^{19} \textrm{cm}^{-2}$, using the relation from \cite{2009MNRAS.400.2050G}.

\subsection{SED Modeling and WD Mass}
\label{sec:WDmass}

\begin{figure}[h]
    \centering
    \includegraphics[scale=0.4]{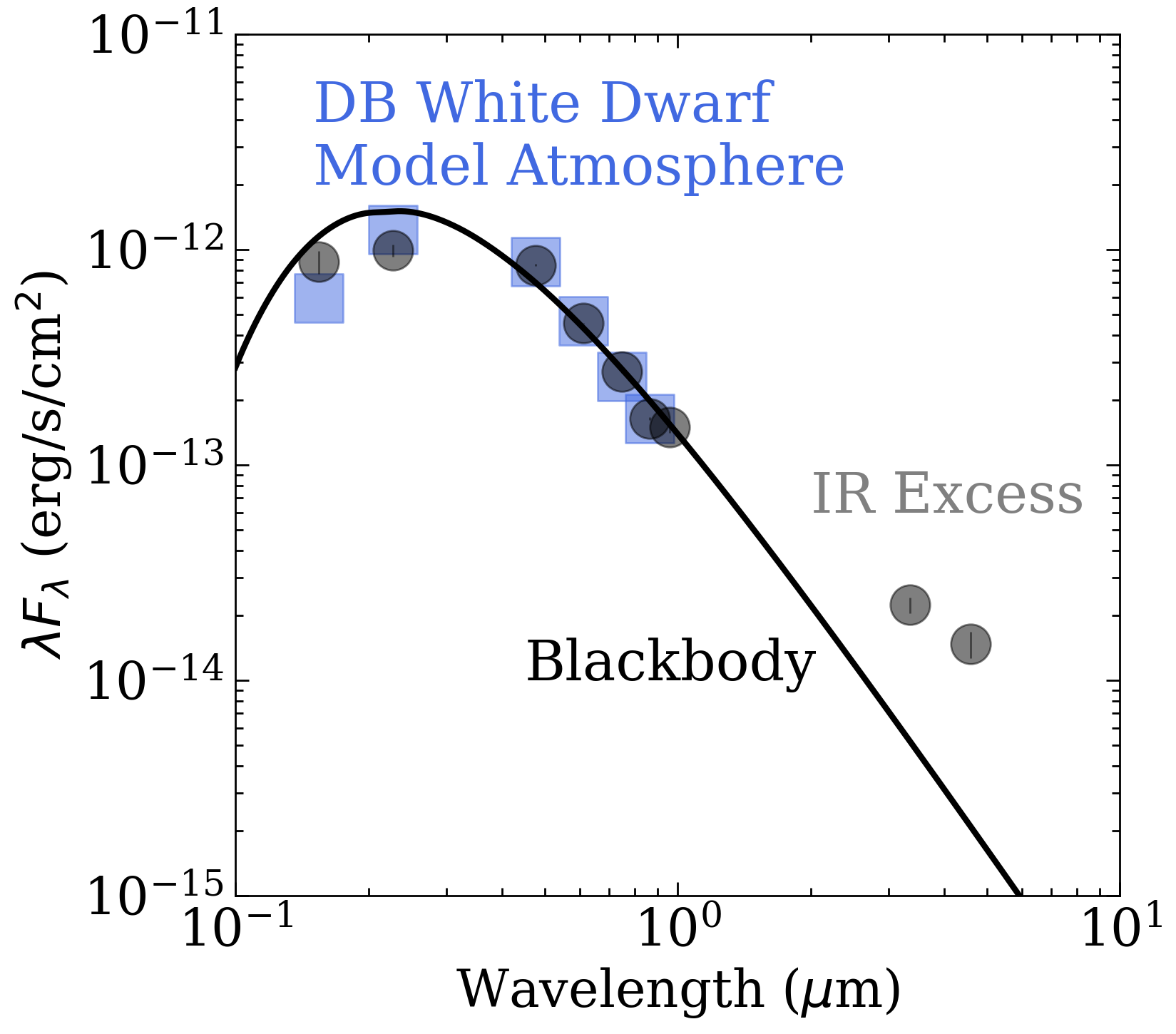}
    \caption{The SED of SRGeJ0453 is well fit by a $T_\textrm{eff}=16,570$ K black body (solid, black line) or $T_\textrm{eff}=12,210$ K DB WD model atmosphere (blue squares) at UV/optical wavelengths. There is an IR contribution from the donor and/or accretion disk. }
    \label{fig:sed}
\end{figure}

We construct the SED of SRGeJ0453 using photometry from GALEX \citep{2005galex}, PanSTARRS \citep{2016panstarrs}, and WISE \citep{2010wise}, compiled from the VizieR\footnote{\url{https://vizier.cds.unistra.fr/viz-bin/VizieR}} and Barbara A. Mikulski Archive for Space Telescopes (MAST)\footnote{\url{https://archive.stsci.edu/}} databases. For the WISE data, we make use of the CatWISE \citep{2020catwise} catalog, which accounts for proper motions and uses stacked WISE images to go deeper than the AllWISE catalog (in which SRGeJ0453 is not present). 

We model the SED, excluding CatWISE data, with a black body corresponding to the accreting WD. Since SRGeJ0453 is a long-period AM CVn ($P_\textrm{orb} \gtrsim 50$ minutes), we assume the accretion disk contribution in the optical wavelengths is negligible due to the low accretion rate. We do not model the bright spot since its contribution is also negligible at optical wavelengths (though this may not be true in the UV). Two other long-period AM CVn systems ZTFJ0220+21 ($P_\textrm{orb} = 53.5$ minutes) and ZTFJ0003+14 ($P_\textrm{orb} = 55.5$ minutes) were found to have the disk and bright spot contribute to the continuum $<$ 8 percent at optical wavelengths ($g$ and $r$ bands) \citep{2022vanroestel}. The light curve of SRGeJ0453 similarly does not show any evidence of a significant disk or bright spot contribution, justifying our approximation.

We perform a Bayesian analysis using the Markov Chain Monte Carlo (MCMC) technique to sample the posterior distribution of the WD radius, WD effective temperature, and extinction, $A_V$. We use an affine invariant sampler as implemented in \texttt{emcee} \citep{2013emcee}. We assume a Gaussian likelihood and set a uniform prior on $R_\textrm{WD}$ (0.005 -- 0.5)$R_\odot$, a uniform prior $T_\textrm{eff}$ (8,000 -- 30,000) K, and a Gaussian prior on the extinction ($E(B-V) = 0.010\pm 0.004$). We run the sampler for 4000 steps, taking half as the burn-in period. The $\chi^2$/dof from the black body approximation is 14.67/5. We determine the WD radius to be (with errors from the 16th and 84th percentiles): $R_\textrm{WD} = 7.39^{+0.09}_{-0.08} \times 10^{-3}R_\odot$. From this radius, we determine the mass to be $M_\textrm{WD} = 0.854^{+0.04}_{-0.05} M_\odot$, following the WD mass-radius relation from \cite{1972shipman}. We chose this WD relation because the accretion rate is so low that it is unlikely to inflate the WD of order more than a few percent (see Section \ref{sec:xray}). We determine the WD effective temperature to be $T_\textrm{eff} = 16,570^{+240}_{-250}$ K. In Figure \ref{fig:sed}, we present the SED of SRGeJ0453 with the WD model resulting from the MCMC analysis. There is clear evidence of an IR excess in WISE W1 and W2 bands. This has been seen in other AM CVns and attributed to the donor and/or accretion disk \citep[e.g.][]{2022vanroestel}. 

In addition to the black body modeling, we also modeled the UV/optical photometry of SRGeJ0453 using DB (pure helium) WD atmosphere models. We used the \texttt{WDPhotTools}\footnote{\url{https://github.com/cylammarco/WDPhotTools}} package \citep{wdphot}, which uses the DB model atmospheres from \cite{2011db_atmos}. We leave $T_\textrm{eff}$ and $M_\textrm{WD}$ as free parameters in this code, which uses the \texttt{emcee} MCMC sampler to explore the parameter space and generate the posterior distribution. The $\chi^2$/dof from the DB atmosphere model approximation is 14.78/5. Using this approach, we determine the WD temperature to be $T_\textrm{eff} = 12,200^{+420}_{-400}$ K. The WD mass is $M_\textrm{WD} = 0.60^{+0.05}_{-0.05} M_\odot$, and also computed is the surface gravity: $\log g = 8.02^{+0.08}_{-0.08}$. With these two values, the WD radius is $R_\textrm{WD} = 1.2^{+0.1}_{-0.1} \times 10^{-2}R_\odot$ (not accounting for any inflation due to accretion heating). Consequently, any parameters computed in this paper that depend on $M_\textrm{WD}$ could be scaled by a factor of 0.71 to account for the difference between the black body and DB atmosphere models.

However, we emphasize that DB model atmospheres still constitute an approximation. Various metals (N, Na, Mg, Ca) are detected in the optical spectrum of SRGeJ0453, all of which have strong resonance lines in the UV that are not modeled in the pure helium atmospheres \citep{2011db_atmos}. We note that both the black body and DB atmosphere models lead to a sub-optimal reduced chi-squared of about 3. This could be explained by neither model actually being a good approximation to the true WD atmosphere in this system. Additionally, it is unclear if there is any contribution from the hot spot or boundary layer in the UV, which DB model atmospheres cannot reproduce. Rather than add the complications associated with assuming a DB model atmosphere, we proceed with all calculations using the black body estimates.

\subsection{ X-ray Properties }
\label{sec:xray}
\begin{table}
\fontsize{10}{15}\selectfont
\centering
\caption{Results of approximation of  X-ray spectrum of SRGeJ0453 by different models. }
\label{tab:Xspectra}
\begin{tabular}{lcc}
\hline                     
\multicolumn{2}{l} { Model: ${\tt tbabs\times (powerlaw) }$ }           \\
\hline
{\it Parameters:}               &                                \\
$N_{\rm H}$($\times 10^{22}$cm$^{2}$)  & $\la0.03$   \\
$\Gamma$          & $0.63^{+0.27}_{-0.25}$     \\
$\rm C-stat$/(d.o.f)            & 11.5/16            \\
\hline     
\multicolumn{2}{l} { Model: ${\tt tbabs\times (mekal) }$ }           \\
\hline
{\it Parameters:}                  &                                \\
$N_{\rm H}$($\times 10^{22}$cm$^{2}$)  & $0.19^{+0.17}_{-0.11}$   \\
$kT_{\rm mekal}$(keV)            & $\ga 3.2$     \\
$\rm C-stat$/(d.o.f)            & 13.4/16            \\
\hline
\hline
$F_{\rm 0.3-2.3\ keV}^{\rm } $  $\rm (\times 10^{-14}\ erg\ cm^{-2}\ s^{-1})$  & $9.0\pm 0.6$   \\
$L_{\rm 0.3-2.3\ keV}^{\rm } $ $\rm (\times 10^{29}\ erg\ s^{-1})$  & $6.2\pm0.4$   \\
\hline
\end{tabular}
\end{table}

\begin{figure}[t]
\centering
\includegraphics[width=0.48\textwidth,clip=true]{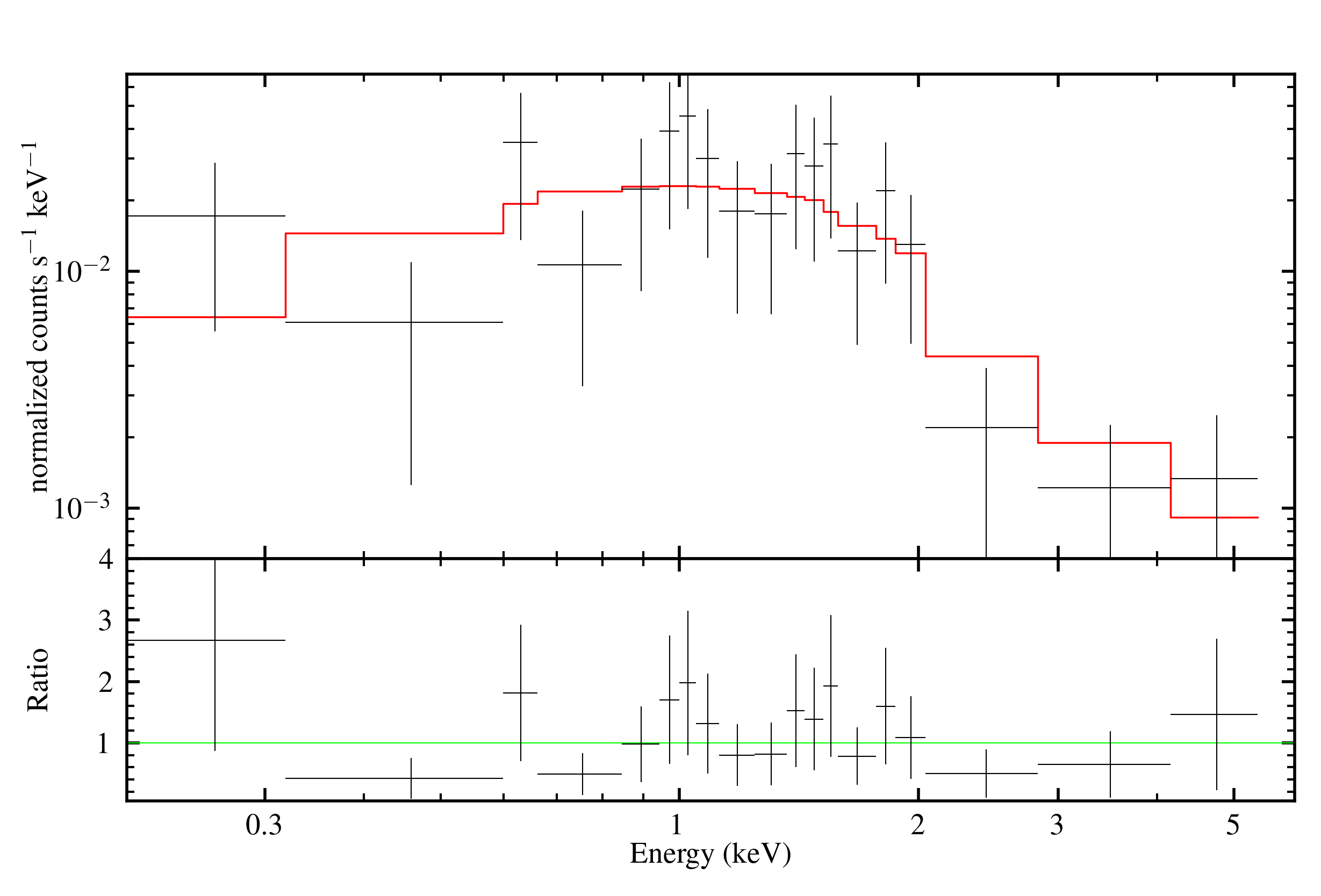}
\caption{The X-ray spectrum of SRGeJ0453 over four SRG/eROSITA all-sky surveys data (top panel). A red line shows the best-fit power-law model from Table \ref{tab:Xspectra}. The bottom panel shows the residuals (ratio of the data divided by the model) in each energy channel. }
\label{fig:Xray_spectrum}
\end{figure}

The X-ray spectrum of SRGeJ0453 is shown in Figure \ref{fig:Xray_spectrum}. The spectrum was obtained by combining four SRG/eROSITA all-sky survey data. We approximated the spectrum of the sources using two models: the power-law model ({\tt powerlaw} in XSPEC) and the optically thin thermal plasma model ({\tt mekal} in XSPEC) with solar abundance. To take into account interstellar absorption, we used the Tubingen-Boulder ISM absorption model ({\tt tbabs} model in XSPEC, \citealt{2000ApJ...542..914W}). We used {\it cflux} task in XSPEC to compute the unabsorbed fluxes. The results of the approximation of the X-ray spectrum of SRGeJ0453 are shown in Table \ref{tab:Xspectra}. The hydrogen column densities estimated from the approximation of the X-ray spectrum are low and agree with the Galactic absorption column density in the direction of SRGeJ0453. Only a lower limit of the palsma temperature $\ga$ 3.2 keV ($1\sigma$ confidence) is estimated from the X-ray spectroscopy.

The absorption-corrected X-ray flux of SRGeJ0453 in the 0.3--2.3 keV energy band is  $(9.0\pm 0.6) \times 10^{-14} \textrm{ erg s}^{-1}\textrm{cm}^{-2}$, computed from the power-law model approximation. Given the well-constrained distance from \textit{Gaia}, the X-ray luminosity is $(6.2\pm 0.4) \times 10^{29} \textrm{ erg s}^{-1}$. 

Assuming half of the gravitational potential energy is radiated away in the boundary layer, as in non-magnetic CVs, the accretion luminosity is:
\begin{align}
    L_\textrm{acc} = \frac{\eta}{2}\frac{GM_\textrm{WD}\dot{M}}{R_\textrm{WD}} 
    \label{eq:acc}
\end{align}
where $M_\textrm{WD}$ and $R_\textrm{WD}$ are the mass and radius of the WD from the results of SED modelling (see Section \ref{sec:WDmass}), and $\eta$ is the radiative efficiency of accretion ($\eta = 1$). We assumed the canonical bremsstrahlung model to compute the bolometric correction (BC) factor for the X-ray luminosity in the 0.3--2.3 keV energy band. For the fixed temperature in the 2--50 keV range, we compute the BC factor in the $\approx 1.8-11.1$ range. Taking this into account, we obtain an accretion rate of $\dot{M} \approx (2-10)\times 10^{-12}\ M_\odot \textrm{ yr}^{-1}$, where uncertainties are caused by the BC factor.

\begin{figure}[t]
\centering
\includegraphics[width=0.5\textwidth,clip=true]{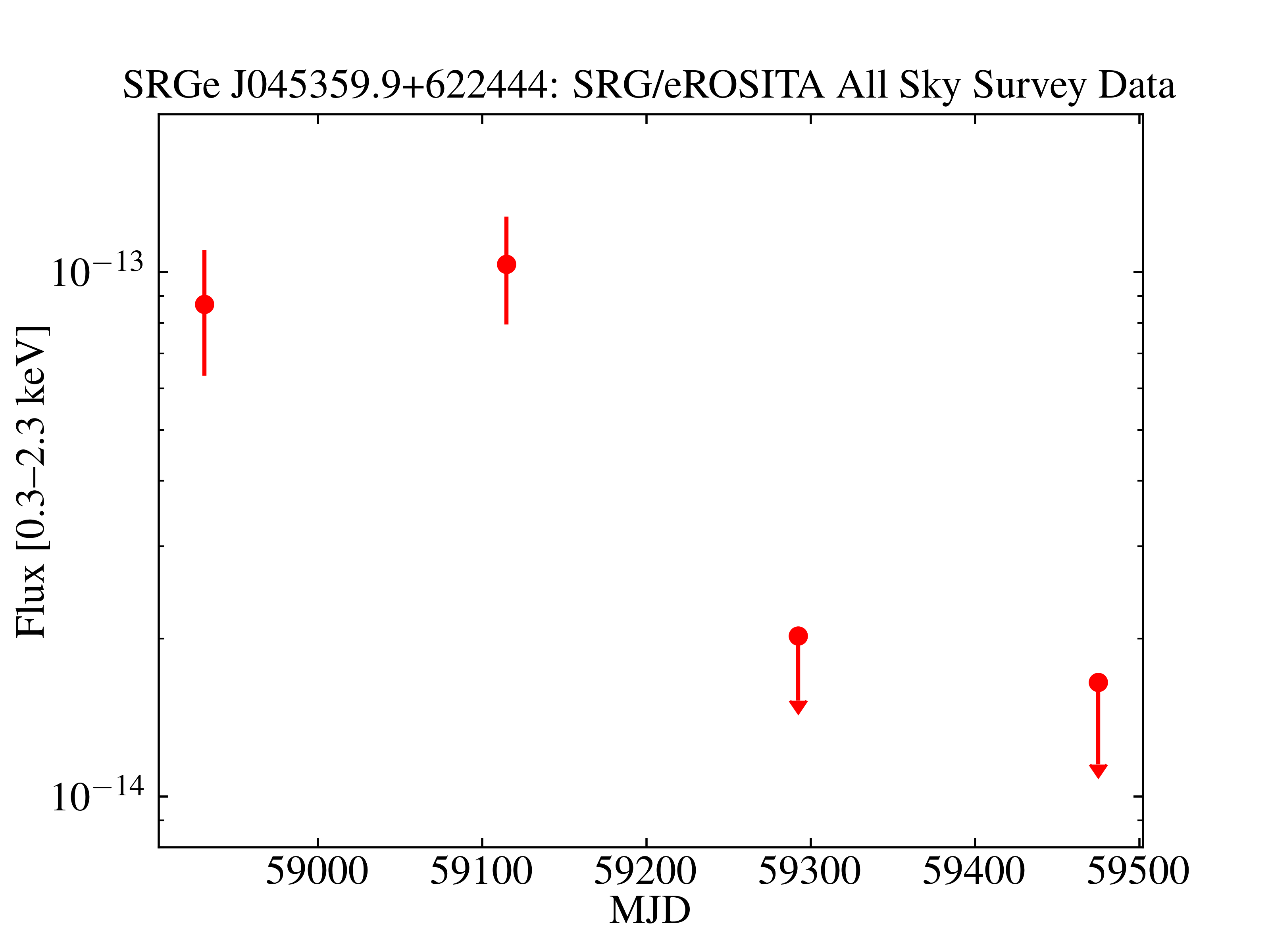}
\caption{The 0.3--2.3 keV X-ray light curve of SRGeJ0453 during four SRG/eROSITA all-sky surveys. Arrows show $\rm 3\sigma$ upper limits for X-ray fluxes.}
\label{fig:Xray_LC}
\end{figure}

Figure \ref{fig:Xray_LC} shows the X-ray light curve of SRGeJ0453 within four sky surveys of SRG/eROSITA. We see the variability in X-rays, where the $3\sigma$ lower limit for the ratio between maximum and minimum fluxes is $\ga$6.3. X-ray variability could correlate with the system's orbital period, but this needs to be investigated by further X-ray and optical follow-up.

We note that the approximation of the X-ray spectrum of SRGeJ0453 by the power-law model gives a photon index of $\Gamma\sim 1$. Previous studies of a dozen classical novae in the Galaxy observed by SRG/eROSITA in quiescence revealed a dichotomy between their photon indexes \citep[e.g.,][]{2021AstL...47..587G}. X-ray spectra of several intermediate polars \citep[magnetic CVs believed to harbor WDs with magnetic field strengths $B\approx1-10$ MG][]{2017mukai} and candidate IPs are approximated by photon index $\Gamma\sim 1$. Non-magnetic systems are approximated by photon index $\Gamma\sim 2$. The hard spectrum of SRGeJ0453 may suggest that it harbors a magnetized WD, however, any meaningful conclusion would require further observations, including polarimetric studies.

\subsection{Phase-Resolved Spectroscopy}
\subsubsection{Systemic Velocity}
\label{sec:sys_vel}

We first constrain the system radial velocity (RV) by identifying He I disk emission lines. We choose several lines that are not blended with other elements (e.g. this eliminates He I 5875.6 \AA\; due a strong blend with 
Na~I). We also exclude lines that are affected by telluric features or show deep central absorption (see Appendix \ref{sec:line_selection}). The remaining He~I lines that fit these criteria are centered at 6678.2 and 7065.2 \AA. We fit two Gaussian profiles to each of the two peaks of the He I emission lines. We then produce RV curves using the center of each of the two Gaussians. We only analyze the middle four orbital phases, since spectra centered at orbital phases between 0.9--1.1 are affected by the donor eclipsing the disk. Spectra centered at these orbital phases, unless centered perfectly at conjunction, will show the donor star preferentially eclipsing a component of the disk that is blue/redshifted (see Appendix \ref{sec:helium_rv}). 

We assume a circular orbit with a RV equation of the form:
\begin{align}
    \textrm{RV} = \gamma +  K_x\sin(2\pi \phi) + K_y \cos(2\pi\phi), 
    \label{eq:radial_velocity}
\end{align}
where $K_x$ and $K_y$ are the two components of the RV, and $\gamma$ is systemic velocity.
We use an MCMC technique to sample the posterior distribution, assuming a uniform prior on $\gamma, K_x,$ and $K_y$ (all ranging between --300 and +300 km/s), and a Gaussian likelihood. We take the standard error of the mean (of the fit to the Gaussians) as the error of each RV. The median value of all parameters and error (16th, 84th quartiles) are in Table \ref{tab:rv_disk}. The weighted average of the systemic velocity of both He~I lines represents the systemic velocity of the binary system as whole: $\gamma_\textrm{binary}=14\pm5$ km/s.

We also present the RV results for the He~II 4685.7~\AA\; line in Table \ref{tab:rv_disk}, which we discuss in more detail in Section \ref{sec:central_spike}.

\begin{table}
	\centering
	\caption{Radial Velocity Measurements of He  Lines
	}
	\label{tab:rv_disk}
	\begin{tabular}{lccc} % four columns, alignment for each
		\hline
		Line (\AA) &   {\parbox{1cm}{\vspace{5pt}\centering $\gamma$ (km/s)}} & {\parbox{1cm}{\vspace{5pt}\centering $K_x$ (km/s)}}  &{\parbox{1cm}{\vspace{5pt}\centering $K_y$ (km/s)}}\\
		\hline
He I\ 6678.2 & $11\pm 4$& $29\pm 5$& $39\pm 7$\\
He I\ 7065.2  & $17\pm 3$& $-29\pm 4$& $58\pm 5$\\
He II\ 4685.7\footnote{This is the ``central spike" line (see Section \ref{sec:central_spike}).} & $192\pm 5$ & $-5 \pm 8$ & $7\pm 6$\\
\hline

	\end{tabular}
\end{table}

\subsubsection{Doppler Tomograms}
Doppler tomography converts phase-resolved spectroscopy into a plot of observed RV and line strength as a function of orbital phase. In other words, spectral line information is converted from time and wavelength space to phase and velocity space. In this process, Doppler tomograms disentangle the contribution of the various CV components (e.g. accretion disk, accretion hot spot, donor star) to a given spectral line; see \cite{doptomography} for a review of the method of Doppler tomography. We use the \texttt{doptomog}\footnote{\url{https://www.saao.ac.za/~ejk/doptomog/main.html}} code developed by \cite{2015kotze} to construct Doppler tomograms shown here.

We use 5 Keck LRIS spectra from a single orbit to create Doppler tomograms for the He~I 5876.5~\AA\;, 
6678.2~\AA\; and 
7065.2~\AA\ lines. We use the systemic RV offset calculated after performing a double Gaussian fit to the emission line throughout the entire orbit (see Table \ref{tab:rv_disk}). However, there is strong blending present in He~I~5876.5~\AA\ and there could be an artifact in one of the spectra of 6678.2~\AA\, so we discuss those Doppler tomograms in Appendix \ref{sec:dop_5876.5}. The He I 7065.2 \AA\; Doppler tomograms, along with trailed spectra, are shown in Figure \ref{fig:doppler0}. We present a Doppler tomogram for the other He I lines in Appendix \ref{sec:dop_5876.5}. 

The He I 7065.2 Doppler tomogram shows that line traces the accretion disk, and a prominent bright spot located approximately 45 degrees ahead of the donor star (second quadrant of the plot). We do not see a significant second bright spot as has been seen in previous studies for other AM CVn systems \citep[e.g.][]{green2019}, but other studies have suggested that the second bright spot may be weak \citep[e.g.][]{2022vanroestel}.

We also create a Doppler tomogram for 
He~II~4685.7~\AA\ (Figure \ref{fig:doppler0}), using the average systemic RV offset of that line. The He II 4685.7 \AA\; tomogram shows neither a disk nor the bright spot seen in the He I tomograms. Instead, the He II tomogram shows clear evidence that this line originates near the center of mass of the system --- the ``central spike" seen in other AM CVn systems. We present a detailed analysis of this feature in Section \ref{sec:central_spike}. 

\begin{figure}
    \centering
    \includegraphics[scale=0.2]{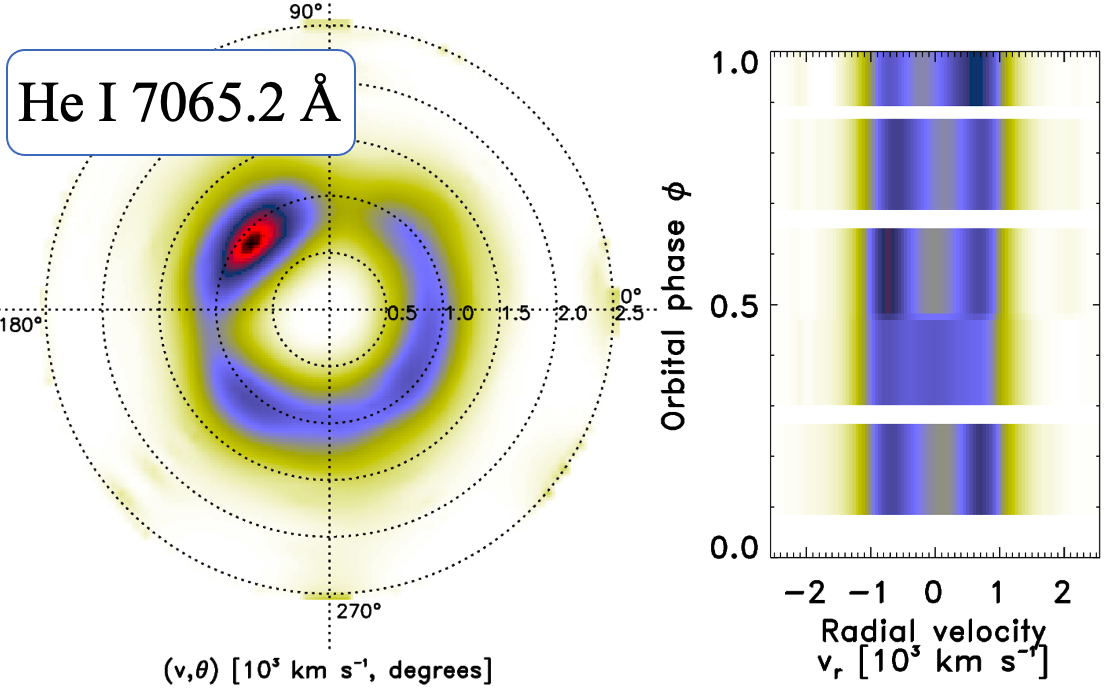}
    \\
    \includegraphics[scale=0.3]{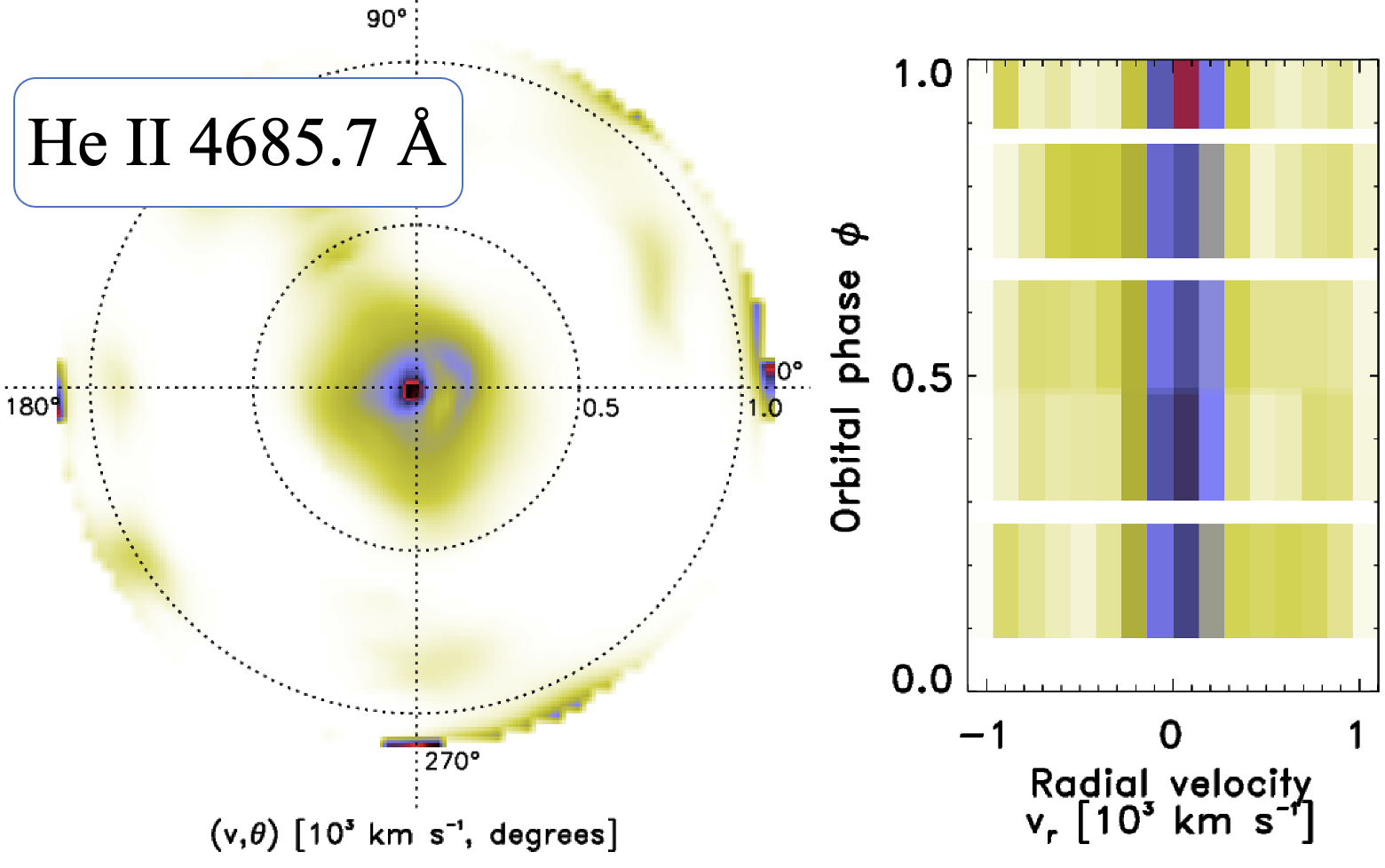}
    \caption{Doppler tomograms and trailed spectra of He~I~7065.2~\AA\; reveal a disk with at least one prominent bright spot (upper panel). The He II 4685.7 \AA\; Doppler tomogram (lower panel) shows the line to originate near the system center of mass, indicating  it is the ``central spike" seen in other AM CVn systems.}
    \label{fig:doppler0}
\end{figure}

\section{Discussion}
\label{sec:discussion}

\subsection{Binary Parameters}
We used two techniques to compute the binary parameters of SRGeJ0453: 1) modelling of the optical light curve, and 2) RV measurements of the He II 4685.7 \AA\ line. The results of these methods are discussed below.

\subsubsection{Light curve Modeling}
\label{sec:LC_modelling}

We modeled the optical light curve using the PHysics Of Eclipsing BinariEs code PHOEBE \citep{2005phoebe1,2016phoebe2}. This code allows to model and fit optical light curves, RV curves, and spectral line profiles of binary systems. PHOEBE constructs 3D models of each star using a triangular mesh and incorporates the physics of irradiation, ellipsoidal modulation, spots, and most relevant to our work, a complete Roche geometry. 

\begin{figure*}
    \centering
    \includegraphics[scale=0.3]{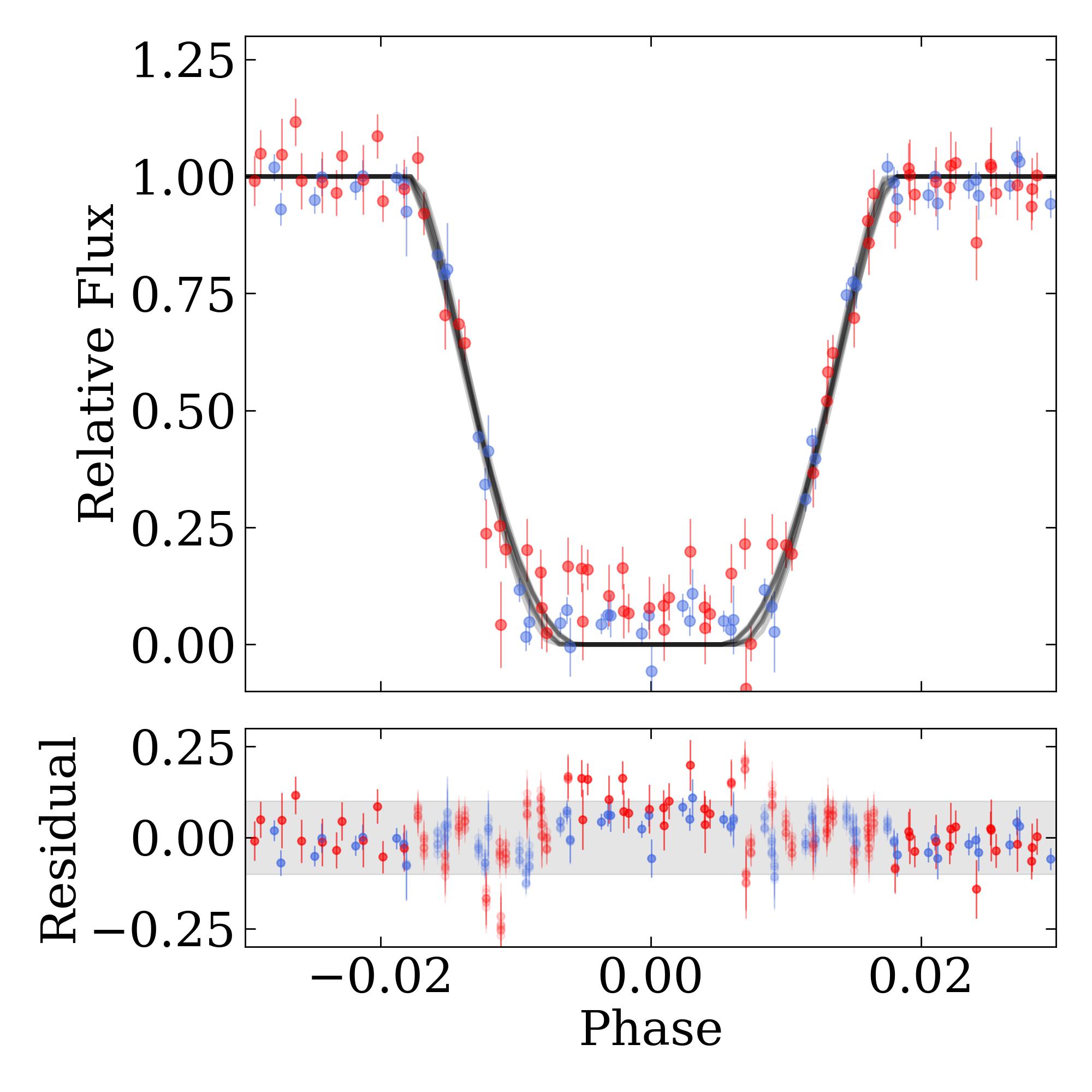}\includegraphics[scale=0.4]{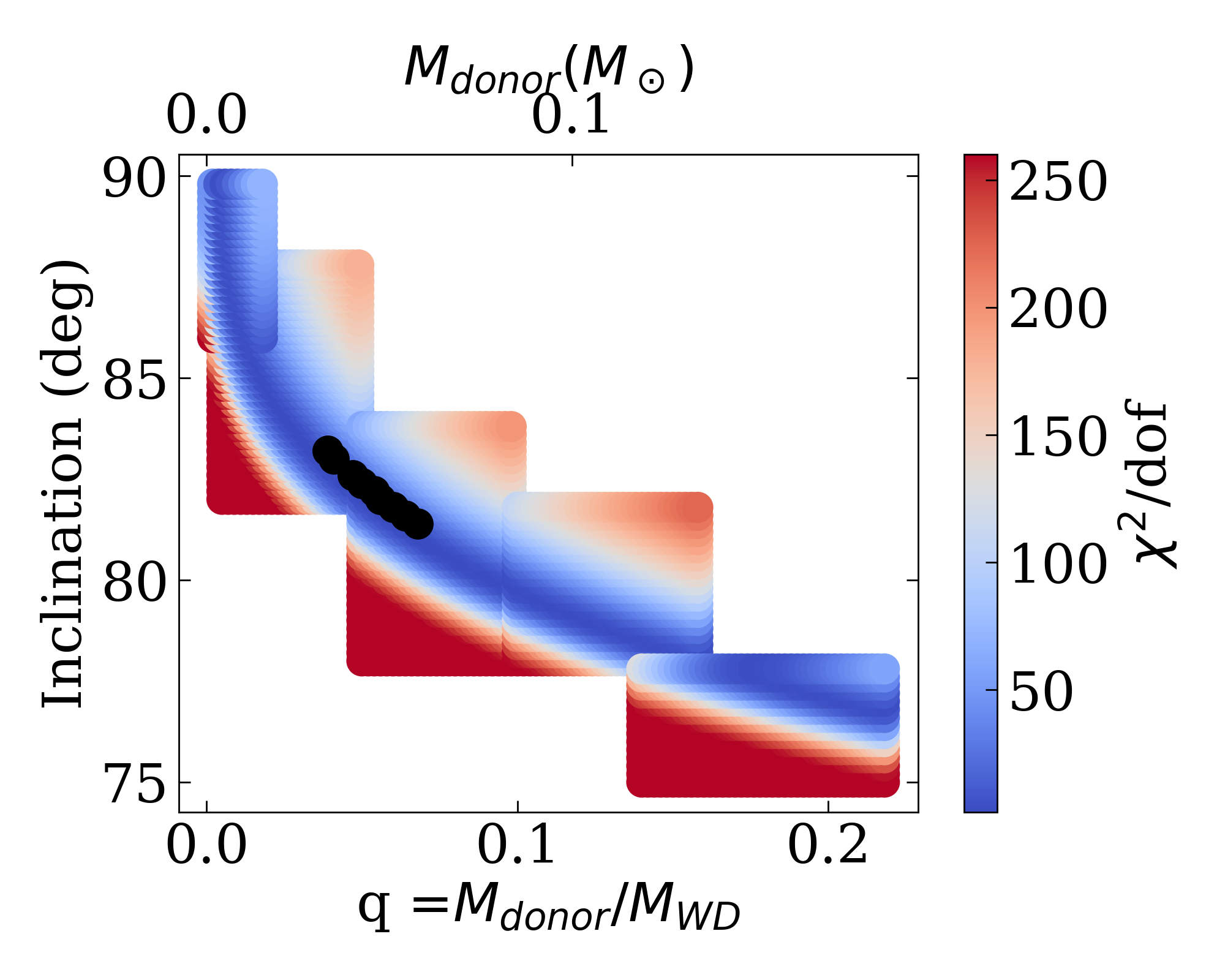}
    \caption{PHOEBE modeling of the eclipsing allows us to constrain model parameters of SRGeJ0453. \textit{Left}: CHIMERA $r$ (red) and $g$ (blue) light curves plotted alongside best-fitting PHOEBE models. \textit{Right}: Goodness-of-fit ($\chi^2$/dof) for all PHOEBE models. In black, we show model parameters where $\chi^2$/dof $<$ 2.3, indicating a good fit to the data. }
    \label{fig:phoebe_chi2}
\end{figure*}

When comparing our PHOEBE models to data, we only made use of the CHIMERA $g$ band light curve, since any emission from the disk and/or bright spot is minimal in that passband. Helium disk emission lines contribute minimally in the CHIMERA $g$ band (equivalent to Sloan $g$: 4000 -- 5500 \AA): The integrated flux over the entire passband (average spectrum) is $\approx 10^{-13}$ erg/s/cm$^2$, while the integrated flux over the most prominent He I emission line, He I 5015.7 \AA, is $\approx 8 \times 10^{-16}$ erg/s/cm$^2$. The contribution of all emission lines in the CHIMERA $g$ passband is therefore $\lesssim 10^{-15}$ erg/s/cm$^2$, around one percent of the continuum flux, which is dominated by the WD.
 
With the above argument in mind, we neglected the emission from the accretion disk and bright spot in our approximation of the $g$ band light curve. Indeed, the contribution of a bright spot does not distort the optical light curve of SRGeJ0453, as expected for non-magnetic CVs \citep[e.g.][]{2019mcallister}. As mentioned in Section \ref{sec:WDmass}, previous results of modelling the $g$ band optical light curve of other long period AM CVns show that the contribution of the accretion disk and bright spot to the integrated emission of the system is usually less than 10 per cent \citep{2022vanroestel}. 

We fixed the WD's mass, radius, and temperature using the SED modelling results (see Section \ref{sec:WDmass}). The optical spectrum and SED of SRGeJ0453 suggests that the donor temperature should be less than 3000 K, since any object at that temperature should be seen in either the spectrum or SED. We fixed the temperature of the donor star at 1500 K. PHOEBE modeling is unaffected by varying the donor star temperature in the 1000--3000 K range or varying the WD temperature between 15,000--18,000 K. The donor star is assumed to fill its Roche lobe, and its radius is computed based on the \citet{1983ApJ...268..368E} approximation: 
\begin{gather}
    \frac{R_L}{a} = \frac{0.49 q^{2/3}}{0.6 q^{2/3} + \ln\left(1 + q^{1/3}\right)},
    \label{eq:eggleton}
\end{gather}
where the semi-major axis of the system $a$ was calculated based on Kepler's law using a best-fit period from section \ref{sec:WDmass} as $a^3 = G(M_1 + M_2)P_\textrm{orb}^2 / 4\pi^2$. The ratio of the donor star mass to the WD mass: $q = M_2/M_1$. Therefore, the two free parameters that we aim to fit are the mass ratio, $q$, and inclination, $i$. 

Outside the eclipse, light curves show flickering at the $<10$ percent level, possibly produced by the accretion disk (see Figures \ref{fig:lc_rtt150}, \ref{fig:RTT150_flc_all} and \ref{fig:CHIMERA}). To compare the PHOEBE model to data, we used only the eclipse in the 0.97--1.03 phase range to minimize the contribution of flickering in parameter estimation. 

We created a grid of PHOEBE models with $0.2^\circ$ steps in $i$ and 0.002 steps in $q$, in the range: $75^\circ \leq i \leq 90^\circ$ and $0.002 \leq q \leq 0.22$. After initially experimenting with a coarser grid over the entire parameter range, we only created PHOEBE models for values of $q$ and $i$ that lead to the best fits to the data due to the high computational cost of creating such a fine grid. In Figure \ref{fig:phoebe_chi2}, we present the resulting fits along with $\chi^2$/dof values. In the left panel of the figure, we present both in-eclipse CHIMERA $g$ and $r$ data (although the fit is only to the $g$ band data) along with the PHOEBE models for which $\chi^2$/dof $<$ 2.3. The residuals show better than 10 percent agreement for the entire range of CHIMERA $g$ band data, while some $r$ band points in eclipse disagree with PHOEBE models at the $\approx$20 percent level. It is unclear if the CHIMERA $r$ band points are due to disk contribution or contamination from the poor seeing that was present during data acquisition.

In the right panel of Figure \ref{fig:phoebe_chi2}, we present the $\chi^2$/dof values for the grid of $q$ vs. $i$ used in our PHOEBE models. The best fitting models ($\chi^2$/dof $<$ 2.3) allow us to obtain the following parameter range: $i \;(^\circ) = 82.5 \pm 1.5$, $q = 0.052 \pm 0.024$. Lower inclinations (and therefore higher values of $q$) are disfavored by the depth of the eclipse, while higher inclinations (and therefore lower values of $q$) are disfavored by the length of WD ingress/egress in the high speed photometry.

The error bars we report $q$ and $i$ reflect the $>99.9\%$ confidence limit. We choose the $\chi^2$/dof $<$ 2.3 range to ensure we do not underestimate our final parameter uncertainties. Visually, all models in this range fit the CHIMERA $g$ band data equally well, which implies that enforcing a lower $\chi^2$/dof value would carry little meaning. This also implies that additional high cadence photometry is needed to place tighter constraints on the binary parameters.  

\subsubsection{``Central Spike": He II 4685.7 \AA}
\label{sec:central_spike}

\begin{figure}
    \centering
    \includegraphics[width=0.45\textwidth]{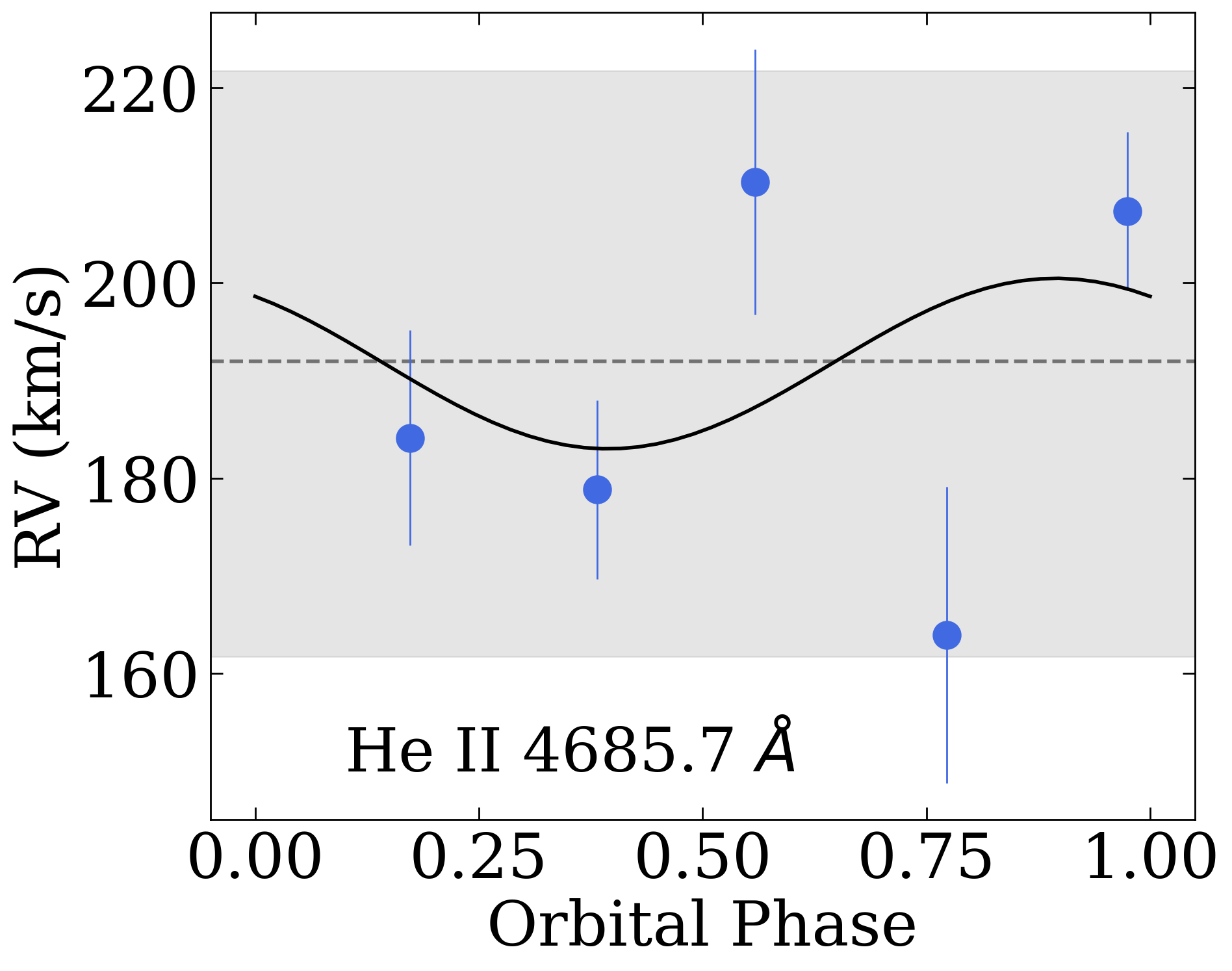}
    \caption{RV measurements of the He II 4686 \AA \;line show a systemic offset (dotted line; $\gamma = 192$ km/s) and an RV amplitude of $K = 9\pm7$ km/s (3$\sigma$ confidence limit shown in grey). This low amplitude suggests that the emission originates near the binary center of mass, and therefore the surface of the WD. For more details, see Section \ref{sec:central_spike}. }
    \label{fig:heII_rv}
\end{figure}

The central spike lines are observed features in the optical spectra of AM CVns. The central-spike lines are thought to originate on the WD surface or close to it \citep[e.g.,][]{1999MNRAS.304..443M,2016MNRAS.457.1828K,green2019}. Therefore, RV measurements for the central spike lines could be used to place limits on the binary parameters of the system.

From the average optical spectrum of SRGeJ0453, we identified the central spike line at He II 4686 \AA \ (see Figure \ref{fig:LRIS}). A similar line was found in the optical spectra of other long-period AM CVns \citep[see, for example,][]{2016MNRAS.457.1828K,green2019,2022vanroestel}. We perform a Gaussian fit to the He II 4686 \AA \; line at every orbital phase. We experiment by trying different sets of initial positions of the line and bounds to find a best-fit line position. The estimated best-fit line positions are independent of initial guesses and bounds. Figure \ref{fig:heII_rv} shows the RV measurements for the central-spike line at He II 4686 \AA. 

The best-fit by Eq. (\ref{eq:radial_velocity}) does not give statistically significant measurements for RVs of the central-spike line, where a systemic velocity of $\gamma_\textrm{sys} = 192 \pm 5$ km/s and RVs of $K_x = -5 \pm 8$ km/s and $K_y = 7 \pm 6$ km/s. The systemic velocity is highly significant, but a similar result might be obtained by fitting the RVs of the central-spike line by a flat line. The chi-squared for a flat line model at the weighted average RV values is $\chi^2$/d.o.f$=11.5/4 = 2.8$, and for the sinusoid function from Eq. (\ref{eq:radial_velocity}) is $\chi^2$/d.o.f$=9.4/2 = 4.7$.  F--test suggests that both models give an acceptable fit (p-value is 0.82) for RV measurements of He II 4686 line.

No significant detection of the RV of the central-spike line at He II 4686 \AA\ is likely due to the low resolution of our data, yet we can still use the amplitude to calculate upper limits ($K =\sqrt{K_x^2+ K_y^2}= 9 \pm 7$ km/s). To constrain the binary parameters, we used a $\rm 3\sigma$ upper limit to the RV amplitude of $K \lesssim 30$ km/s. Assuming that the central-spike line is produced at the WD surface, and its RV is equal to the RV of the WD ($K\approx K_1$), we compute the binary mass function as:
\begin{align}
     \frac{P_\textrm{orb}K_1^3}{2\pi G} =
     \frac{(M_2\sin i)^3}{(M_1 + M_2)^2} = \frac{M_1(q\sin i)^3}{(1 + q)^2},
    \label{eq:K1_WD}
\end{align}
where the mass of the WD is equal to $\rm M_1= 0.85\ M_{\odot}$ (see Section \ref{sec:WDmass}), the orbital period is $\rm P_{orb}= 55.08$ minutes (see Section \ref{sec:period}), and the inclination angle is $i\approx 82\degr$ (see Section \ref{sec:LC_modelling}). Using Eq. (\ref{eq:K1_WD}), we compute $q < 0.052$, which allows us to place limits on both the donor mass and radius:  $ M_2<0.044\ M_{\odot}$ and $R_2\lesssim 0.078\ R_{\odot}$, assuming a Roche-lobe filling star.

The systemic velocity of the He II 4686 line shows the offset from the \textit{system} velocity of the binary ($\gamma_\textrm{binary}=14\pm 5$ km/s) by $\gamma_\textrm{He II 4686} = 178\pm 7$ km/s. A similar excess of the systematic velocities was found in other AM CVns \citep[e.g.][]{green2019}. We emphasize that this systemic redshift is present at \textit{all} orbital phases, ruling out orbital motion as a possible source. This leaves only the gravitational redshift of the WD as a source of the systemic redshift. However, given our measured WD mass and radius, we would only expect a gravitational redshift of $v = GM_\textrm{WD}/(c R_\textrm{WD} )\approx 72$ km/s, which is substantially lower than our measured value. A WD mass $\approx1.1M_\odot$  is required to obtain a gravitational redshift of $\approx180$ km/s, but a WD of this mass seems unlikely given our SED analysis (Section \ref{sec:WDmass}). Further investigation is needed to confirm the large systemic redshift observed in the He II 4685.7 \AA\; line and determine its origin.

\begin{table*}
\caption{Summary of system parameters of SRGeJ0453.
	\label{tab:params}}
\renewcommand\arraystretch{1.13}
\centering
	\begin{tabular}{llc} % four columns, alignment for each
		\hline
		Parameter & Value & Origin\\
		\hline
Distance, $d$ (pc)& $239^{+11}_{-8}$ & Gaia Parallax\\
Orbital period, $P_\textrm{orb}$ (min) & $55.0802 \pm 0.0003$  & Optical Photometry\\
\hline
Extinction, $A_V$ & $0.03 \pm 0.01$ & SED Fit\\
Accretor mass, M$_\textrm{WD}$ ($M_\odot$) & $0.85^{+0.04}_{-0.05}$ & SED Fit\\
Accretor temperature, T$_\textrm{eff,WD}$ (K) & $16,570^{+240}_{-250}$ & SED Fit\\
\hline
Inclination, $i$ ($^\circ$) & $82.5\pm 1.5$& PHOEBE Model\\
Mass ratio, $q$ & $0.052 \pm 0.024$ & PHOEBE Model\\
 & $\la0.052$ & Spectroscopy\\
\hline
 Accretion rate, $\dot{M}$ ($M_\odot \textrm{ yr}^{-1}$) & $\approx (2-10)\times 10^{-12}$& X-ray\\
 \hline
Donor mass, M$_\textrm{donor}$ ($M_\odot$)  & $0.044 \pm 0.020$ & SED Fit + PHOEBE\\
 & $\la0.044$ & SED Fit + Spectroscopy\\
Donor radius, R$_\textrm{donor}$ ($R_\odot$) & $0.078 \pm0.012$ & SED Fit + PHOEBE\\
 & $\la0.078$ & SED Fit + Spectroscopy\\
\hline
	\end{tabular}

\end{table*}

\subsection{Evolutionary History}
\label{sec:evolution}

We discuss the evolutionary history of the system using lines of H, C, N, and O. We do not detect H in the optical spectrum of SRGeJ0453. In the \citet{1985SvAL...11...52T} evolved donor channel model, based on the empirical \citet{1981A&A...100L...7V} magnetic braking picture, the surface mass abundance of hydrogen of systems with $P_\textrm{orb}\approx$55 minutes never drops below several 0.01 in a Hubble time \citep[see Table A1 in ][]{2003MNRAS.340.1214P}. In this scenario, H should be detected by modern techniques (T. Marsh, private communication). However, in the $\alpha -\Omega$\ dynamo picture for magnetic braking in the evolved CV model \citep{2023sarkarevolvedcv}, and in the modeling done in \cite{2021elms}, the hydrogen mass fraction in long-period AM CVns may be present at a non-detectable level. Thus, our non-detection of H in SRGeJ0453 cannot rule out the evolved donor channel.

\cite{2010nelemans} calculated abundances for systems evolving through all three postulated AM CVn formation channels. A useful diagnostic from that work is the N/O ratio, which in the evolved CV channel, is N/O$\approx$1 \citep[see Figure 6 of][]{2010nelemans}. In the He WD and He star channels, it is possible to have N/O$>$1. \citep[see Figure 11 of][]{2010nelemans}. Since there is a clear non-detection of oxygen and a strong detection of nitrogen, we can discard the evolved CV channel for SRGeJ0453. This leaves us with the He star and He WD channels as possibilities.

The N/C ratio may be also informative in distinguishing between formation channels \citep{2010nelemans}. For the He star channel, \citet{2010nelemans}, using the ``standard" magnetic braking model and \citet{2023sarkarHEstar}, using the double-dynamo model, predict N/C$\approx$5 and 120--200, respectively, while for the He WD and evolved CV channels N/C$\gtrsim$ 100 in both models. However, based on the equivalent widths of the N and C lines, we are able only to estimate for SRGeJ0453 a lower limit of N/C$\gtrsim$1, thus allowing for both He WD and He star channels.
 
It is not surprising that it is difficult to distinguish between the He star and He WD formation channels, since at long orbital periods, donors in systems forming from either channel have nearly identical entropies. This is due to the thermal timescale becoming nearly equal to the mass-transfer timescale, leading to the donor in the He-star channel becoming nearly degenerate \citep[e.g.][]{2010solheim}. 

We conclude that even with precise abundance or donor mass and radius estimates, the uncertainties in current evolutionary models of any formation channel are large enough to accommodate more than one channel for some AM CVn systems \citep[e.g.][]{2022vanroestel}.

\subsection{Comparison to Other AM CVns and Future Work}
SRGeJ0453 is the ninth published eclipsing AM CVn --- YZ LMi/SDSSJ0926+3624 was the first \citep{yzlmi}, Gaia 14aae the second \citep{gaia14aae_discovery}, ZTFJ1905+3134 the third \citep{2020burdge}, and the fourth through eighth were discovered through a dedicated search for eclipsing AM CVns in ZTF \citep{2022vanroestel}. Like \textit{in all other} eclipsing AM CVns with WD mass measurements, the WD in SRGeJ0453 has a mass equal to or greater than that of single WDs \citep[$\overline{M}_\textrm{WD}\approx 0.6M_\odot$; ][]{2007kepler}, regardless of the black body or DB model atmosphere approximation. This presents further evidence that the mean mass of the WD in AM CVns is greater than that of single WDs, as has been found for CVs (see \citet{2022pala} and references therein). 

\begin{figure}
    \centering
    \includegraphics[width=0.5\textwidth]{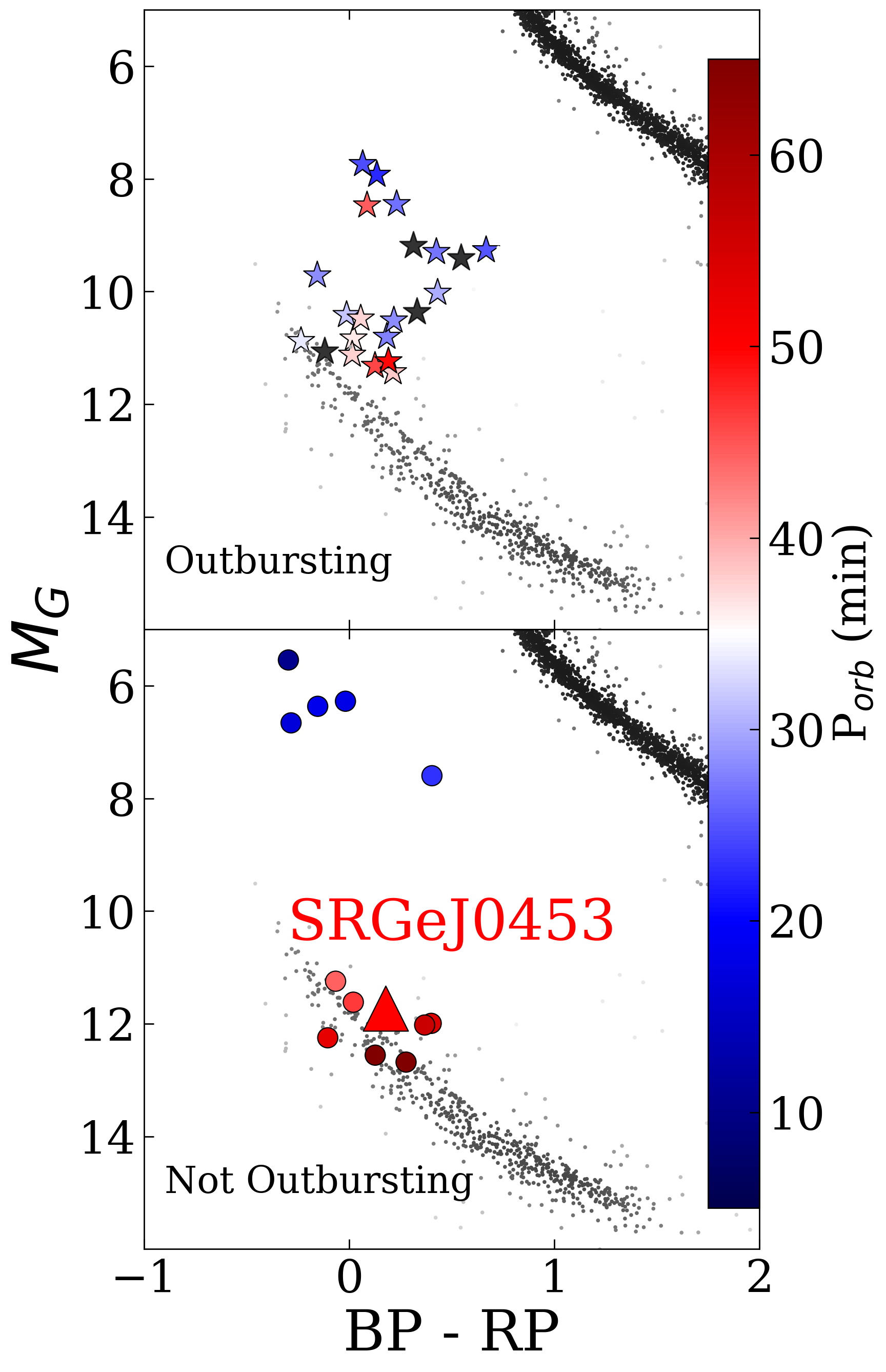}
    \caption{Position of SRGeJ0453 in the 100 pc Gaia Hertzsprung-Russell diagram alongside previously known AM CVn systems with a significant Gaia parallax ($\texttt{parallax\_over\_error} > 3$). The non-outbursting population occupies distinct portions of phase space. }
    \label{fig:gaia_cmd}
\end{figure}

We plot the location of SRGeJ0453 in a 100 pc Gaia Hertzsprung-Russell diagram \citep{gaiadr3} along with other other known AM CVn systems assembled from the catalogs by \cite{2018ramsay} and \cite{2022vanroestel} in 
Figure~\ref{fig:gaia_cmd}. We only plot systems for which a significant Gaia parallax is known ($\texttt{parallax\_over\_error} > 3$). Only systems for which a period is known are then color-coded. We make a distinction between systems that show frequent \textit{optical} outbursts and those that spend the vast majority of their time in optical quiescence. 

Figure \ref{fig:gaia_cmd} shows the progression of AM CVn systems throughout their accreting lifetimes. They start as non-outbursting, short period ($P_\textrm{orb} = 5-10$ minutes), whose high luminosity makes them easy to discover. As they evolve to moderate periods ($P_\textrm{orb} = 20-50$ minutes), they become less luminous, but undergo frequent outbursts which can be discovered by all-sky optical surveys. Finally, AM CVns become long-period ($P_\textrm{orb} \gtrsim 50$ minutes) systems such as SRGeJ0453, which blend in with the WD track. 

We have shown that our multiwavelength analysis has led to the discovery of a long period AM CVn system, due to it being an X-ray source in the SRG/eROSITA sky. With its observed flux of $9 \times 10^{-14}\ \textrm{erg s}^{-1}\textrm{cm}^{-2}$ in the 0.3--2.3 keV energy band, SRGeJ0453 is not present in the Second ROSAT All Sky Survey Source Catalog, which had a flux limit of $\sim 2\times 10^{-13} \textrm{erg s}^{-1}\textrm{cm}^{-2}$ \citep[2RXS;][]{2016bollerA&A...588A.103B}. Furthermore, this system was missed in a search for eclipsing AM CVn systems in the ZTF database of optical photometry due to the requirement that several in-eclipse points be present in the light curve \citep{2022vanroestel}. 

Finally, we note that while many AM CVn systems emit GWs that will be detectable by \textit{LISA}, it is unlikely that SRGeJ0453 will be detectable in the initial 4 year long campaign. We use \texttt{LEGWORK} \citep{LEGWORK_joss, LEGWORK_apjs}, a package designed to calculate \textit{LISA} signal-to-noise ratios for GWs emitted from inspiraling binary systems. We estimate that SRGeJ0453 will be observed with a signal-to-noise ratio $\approx$ 1 in the initial 4 year long campaign.

\section{Summary and Conclusion}
\label{sec:summary}

We have presented the first discovery from a joint SRG/eROSITA and ZTF program to search for CVs and related objects in the Milky Way. SRGeJ0453 is an eclipsing, long period AM CVn system which was identified through its high X-ray to optical flux ratio ($F_X/F_\textrm{opt} \approx 0.12$). Our results are summarised as follows:

\begin{itemize}
 \item[--] The optical spectrum of SRGeJ0453 shows common features for AM CVn systems: a blue continuum with prominent He lines and an absence of H lines. All He I lines have a narrow central absorption commonly seen in nearly edge-on CVs and AM CVns among them. 
 He~II~4685.7~\AA\ is only seen as a narrow emission feature (see Figure~\ref{fig:LRIS}).

 \item[--] The optical light curves from observations done with RTT-150/TFOSC, ZTF, and P200/CHIMERA show deep eclipses ($\approx3^m$). Low amplitude ($\approx0.1-0.3^m$) flickering, possibly caused by an accretion disk, is also seen (see Figures \ref{fig:ztf_lc}--\ref{fig:CHIMERA}).  High-speed photometry allows us to tightly constrain the orbital period to be $P_\textrm{orb} = 55.0802 \pm 0.0003$ minutes.

 \item[--] Using Keck I/LRIS spectra, we created Doppler tomograms for several prominent lines (Figure \ref{fig:doppler}). 
 The He~I~7065.2~\AA\ Doppler tomogram shows a bright spot located $\approx$45 degrees ahead of the donor star, but we see at most weak evidence for a second bright spot in other helium lines. The He~II~4685.7~\AA\ tomogram indicates that this line traces the ``central spike" seen in other AM CVn systems. 

 \item[--] To estimate the binary parameters, we modelled the optical light curve using PHOEBE and used the RV measurements of the central spike line at He II 4685.7 \AA. The computed binary parameters are presented in Table \ref{tab:params}.  

 \item[--] SRGeJ0453 has an X-ray luminosity $\approx 6.2\times 10^{29}\ \textrm{erg s}^{-1}$ in the 0.3--2.3 keV energy band. The X-ray spectrum of the source can be approximated by a power law model with a photon index of $\Gamma\sim 1$, while the optically thin thermal plasma model gives a lower limit on the plasma temperature of $\gtrsim 3.2$ keV ($1\sigma$ confidence, see Table \ref{tab:Xspectra}). Intermediate polars (magnetic CVs with $B\approx 1-10$ MG) observed with SRG/eROSITA show similarly hard photon indexes. This could tentatively suggest a possible magnetic nature of SRGeJ0453, but such a hypothesis requires further investigation. 

 \item[--] The lower limit of abundance ratio of N/C $\gtrsim$1 and the absence of oxygen lines suggests that the He WD and He star evolutionary channels could both be possible for SRGeJ0453 (Section \ref{sec:evolution}).
\end{itemize}

A multi-wavelength campaign such as our combination of X-ray and optical information enables the possibility to efficiently search for and discover AM CVns. This method is particularly useful to identify systems that remain quiescent without any outburst activity. This work is a pilot study, and more systems discovered from a joint SRG/eROSITA and ZTF program to identify CVs will be presented in the near future. 

\section{Acknowledgements}
This work is based on observations with eROSITA telescope onboard SRG observatory. The SRG observatory was built by Roskosmos in the interests of the Russian Academy of Sciences represented by its Space Research Institute (IKI) in the framework of the Russian Federal Space Program, with the participation of the Deutsches Zentrum für Luft- und Raumfahrt (DLR). The SRG/eROSITA X-ray telescope was built by a consortium of German Institutes led by MPE, and supported by DLR. The SRG spacecraft was designed, built, launched and is operated by the Lavochkin Association and its subcontractors. The science data are downlinked via the Deep Space Network Antennae in Bear Lakes, Ussurijsk, and Baykonur, funded by Roskosmos. The eROSITA data used in this work were processed using the eSASS software system developed by the German eROSITA consortium and proprietary data reduction and analysis software developed by the Russian eROSITA Consortium.

Based on observations obtained with the Samuel Oschin Telescope 48-inch and the 60-inch Telescope at the Palomar Observatory as part of the Zwicky Transient Facility project. ZTF is supported by the National Science Foundation under Grants No. AST-1440341 and AST-2034437 and a collaboration including current partners Caltech, IPAC, the Weizmann Institute of Science, the Oskar Klein Center at Stockholm University, the University of Maryland, Deutsches Elektronen-Synchrotron and Humboldt University, the TANGO Consortium of Taiwan, the University of Wisconsin at Milwaukee, Trinity College Dublin, Lawrence Livermore National Laboratories, IN2P3, University of Warwick, Ruhr University Bochum, Northwestern University and former partners the University of Washington, Los Alamos National Laboratories, and Lawrence Berkeley National Laboratories. Operations are conducted by COO, IPAC, and UW. The ZTF forced-photometry service was funded under the Heising-Simons Foundation grant \#12540303 (PI: Graham).

We are grateful to the staffs of the Palomar and Keck Observatories for their work in help us carry out our observations. We thank T\"UB\.{I}TAK, the Space Research Institute of the Russian Academy of Sciences, the Kazan Federal University, and the Academy of Sciences of Tatarstan for their partial support in using RTT-150 (Russian - Turkish 1.5-m telescope in Antalya).

This work has made use of data from the European Space Agency (ESA) mission
{\it Gaia} (\url{https://www.cosmos.esa.int/gaia}), processed by the {\it Gaia}
Data Processing and Analysis Consortium (DPAC,
\url{https://www.cosmos.esa.int/web/gaia/dpac/consortium}). Funding for the DPAC
has been provided by national institutions, in particular the institutions
participating in the {\it Gaia} Multilateral Agreement.

This research made use of matplotlib, a Python library for publication quality graphics \citep{Hunter2007}; NumPy \citep{Harris2020}; Astroquery \citep{2019AJ....157...98G}; Astropy, a community-developed core Python package for Astronomy \citep{2018AJ....156..123A, 2013A&A...558A..33A}; and the VizieR catalogue access tool, CDS, Strasbourg, France. The authors wish to thank E. Kotze for making his Doppler tomography code, \texttt{doptomog}, public \citep{2015kotze}.

ACR acknowledges support from the National Academies of Science via a Ford Foundation Predoctoral Fellowship. IG acknowledges support from Kazan Federal University. The work of IB, MG, IKh, AS, PM  supported by the RSF grant N 23-12-00292.

\bibliography{Eclipsing_AM_CVn}{}
\bibliographystyle{aasjournal}

\appendix

\section{Period estimation using RTT-150 data}
\label{app:rtt150_period}

\begin{figure*}[t]
\begin{center}
{
\includegraphics[width=0.6\linewidth,clip=true]{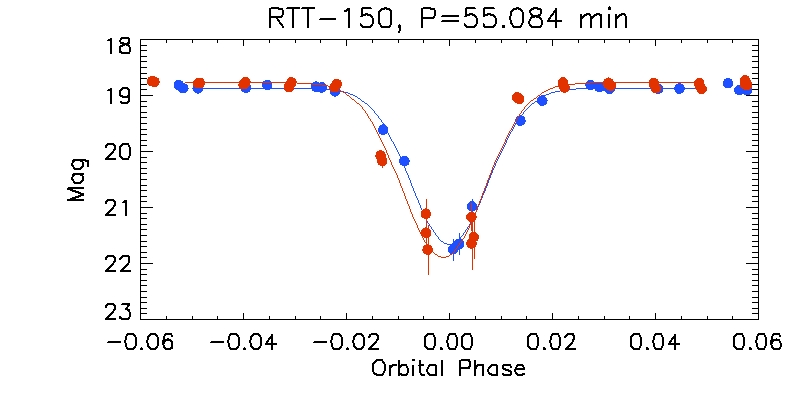}
}
\caption[] { Period determination with two epochs of RTT-150 data comparison. $\it{Blue~circles}$ corresponds to January 19, 2023 data, $\it{red~circles}$ -- January 20, 2023 data with Gaussian fit as solid lines.
}
\label{fig:RTT150_flc}
\end{center}
\end{figure*}

To calculate RTT-150 optical photometry, five stars with good signal-to-noise ratio were used to estimate the position and perform photometric calibration of SRGeJ0453 to the Gaia $G$ band. Fluxes were computed using a 2$\arcsec$ aperture, and the accuracy of RTT-150 differential photometry is 0.02 mag.

A combined approach was used to determine the period of SRGeJ0453. Based on the frequency analysis, a preliminary period was determined and refined by comparing two epochs of observations on different nights. Even though the eclipse profiles for each night in the RTT-150 data are highly smoothed, the two observational epochs differ by about 25 cycles. Thus, the available cycles of eclipse coverage are sufficient to estimate the eclipse center and, therefore, to correct the period itself.

For the frequency analysis, we used the Stellingwerf method implemented in the ISDA code \citep{1978ApJ...224..953S}. This method is one of many variations of the PDM (Phase Dispersion Minimization) method. The period that gives the smallest possible dispersion of the obtained light curve is chosen in the PDM method. This is achieved by minimizing the sum of the squares of ordinate differences from one data point to the next. The periods giving the smallest sum are taken as true periods.

The following algorithm was used to correct the obtained period: (i) The phase curve with the given period is constructed for each night separately; (ii) A phase interval of $\pm0.05$ around the phase, corresponding to the maximum magnitude value, is determined. A Gaussian with a constant fitting of the eclipse region is constructed separately for each night; (iii) The difference modulus between the displacement parameters of the two Gaussians is determined; (iv) Finally, the period corresponding to the minimum of the difference modulus is the best calculated period. The period estimation error is computed as the square root of the sum of the squares of the mean square errors of the Gaussian displacement parameters.

Newton's method finds the minimum of the displacement modulus, dividing the interval in half, where the interval bounds are taken from the preliminary estimation of the period by the Stellingwerf method. The computed period is $55.084 \pm 0.015$ min.

\section{Selection of He I Lines}
\label{sec:line_selection}

\begin{figure}
     \centering
     \includegraphics[scale=0.3]{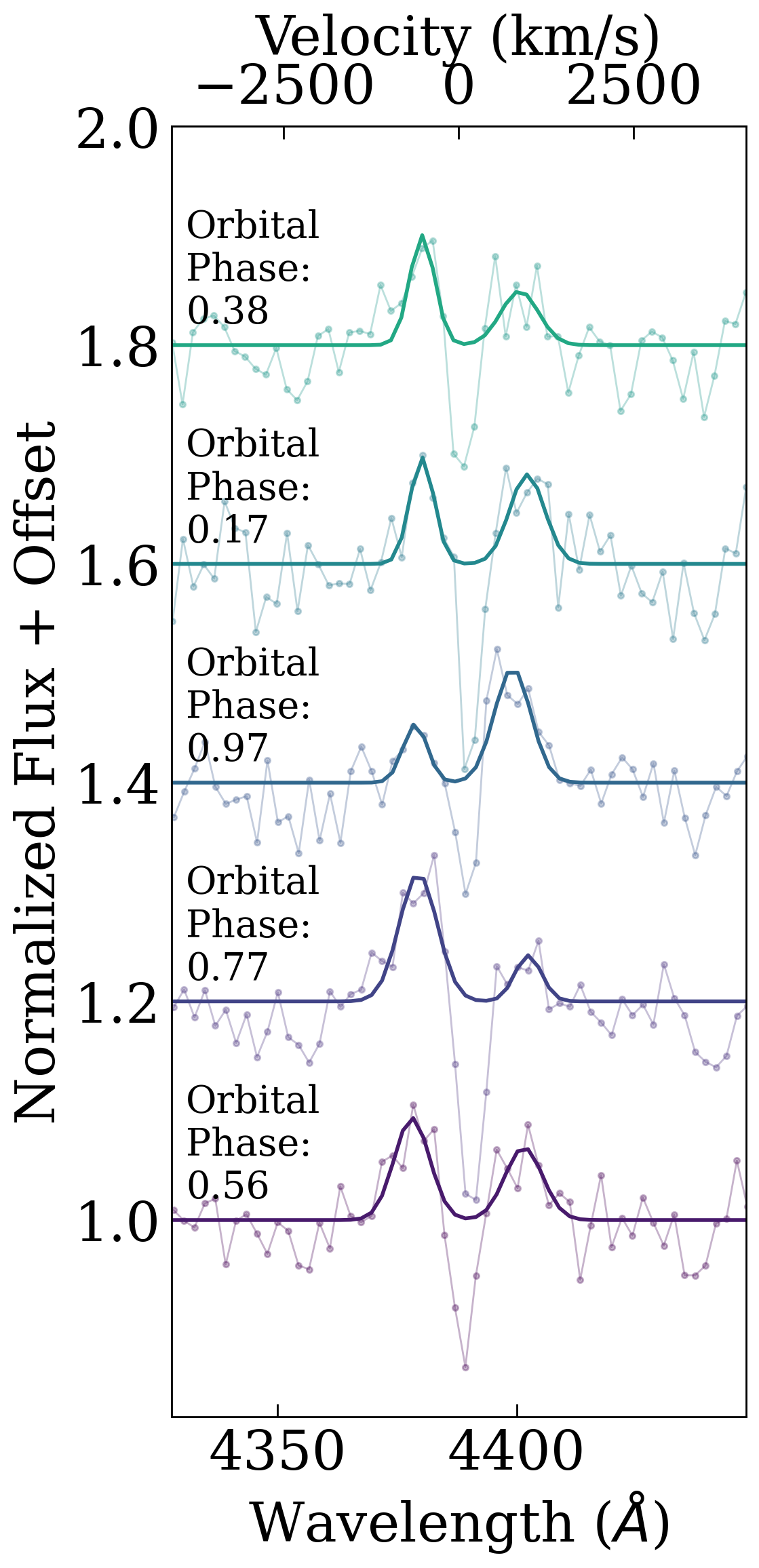}\includegraphics[scale=0.4]{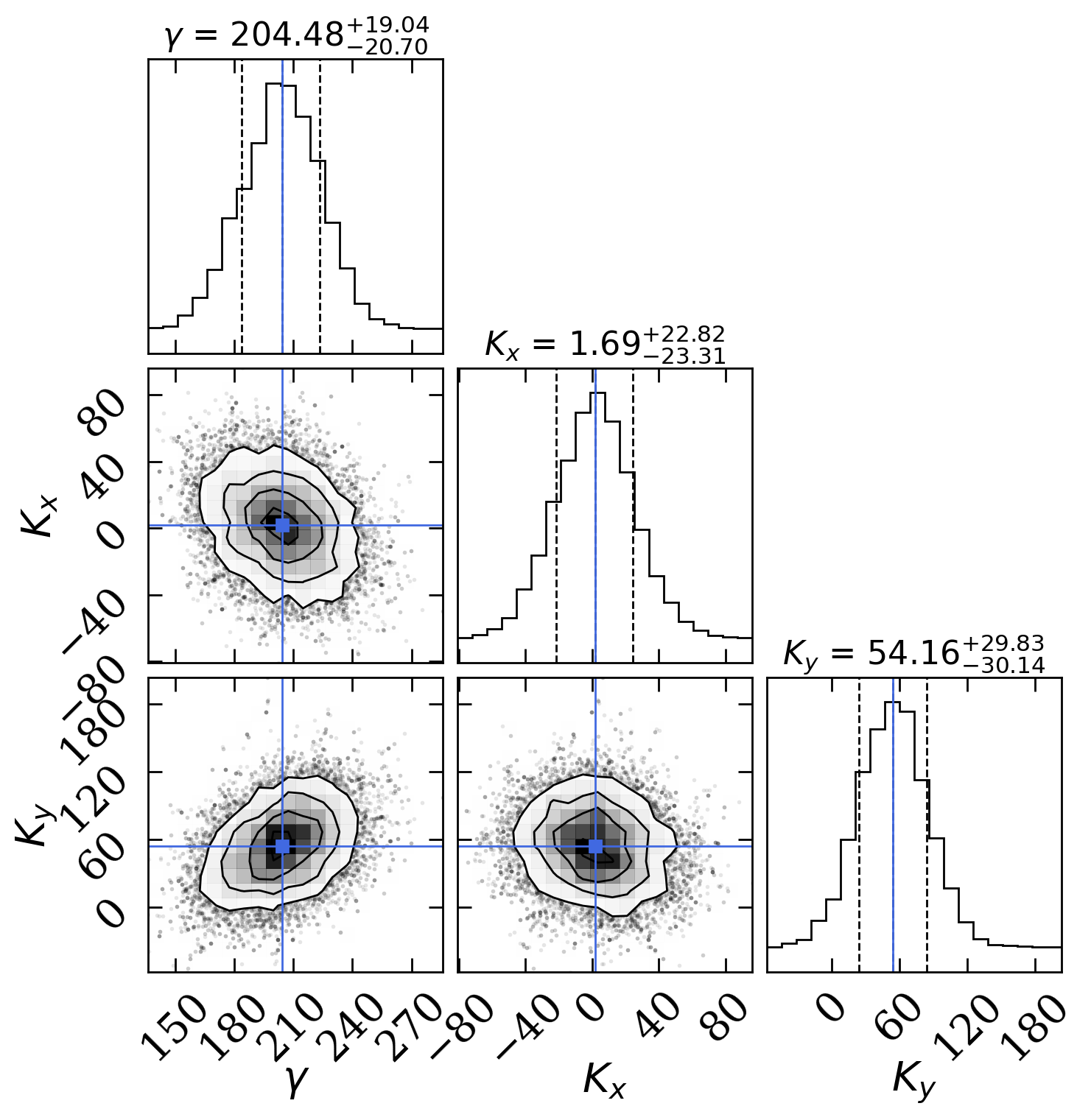}\includegraphics[scale=0.25]{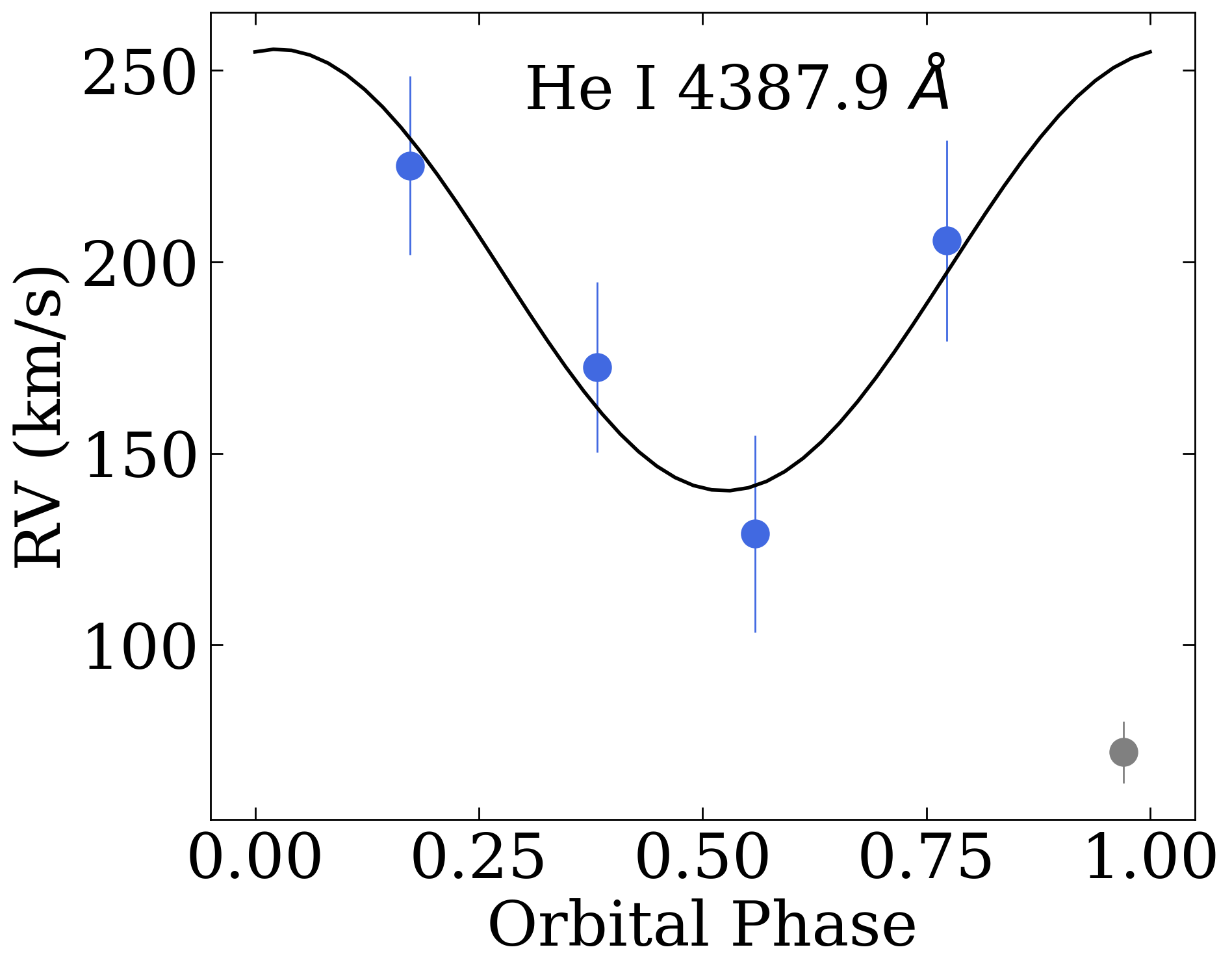}\\
     \caption{Phase-resolved spectra of He I 4387.9 \AA\; with resulting MCMC parameter estimates of radial velocities. The He I 4387.9 \AA\; line is poorly fit due to the strong central absorption feature, which affects the final parameter estimates. Stronger He I \AA\; lines, however, are not affected by low signal-to-noise or strong absorption (see Figure \ref{fig:helium_i_lines_good}).}
     \label{fig:helium_i_lines}
 \end{figure}

We elaborate on why we omit the He I 4387.9 \AA\; and 4921.9 \AA\; lines in the analysis presented in Section \ref{sec:sys_vel}. In Figure \ref{fig:helium_i_lines}, we present phase resolved spectra of the He I 4387.9 \AA\; line, together with the best-fit made up of two Gaussians. We also show MCMC corner plots to show the constraints on $\gamma, K_x, \textrm{and } K_y$, and the resulting RV curves. We show that the He I 4387.9 \AA\; line is affected by both low signal above the continuum and deep central absorption. This absorption seems to be overall redshifted with respect to line center, and has the effect of redshifting the systemic velocity of the line. We attempted to use another fitting procedure where the widths of two Gaussians are held equal, and the central absorption core is omitted from the fit \citep[e.g.][]{green2019}. However, this leaves too few points to fit since the peak of the He I 4387.9 \AA\; line is less than 10\% above the continuum, which results in a poor fit. Exactly the same feature is present in the He I 4921.9 \AA\; line. 

This effect is not seen in the 6678.2 \AA\; and 7065.2 \AA\; lines, which are much stronger (see Figure \ref{fig:helium_i_lines_good}). In that case, the lines are more than 50\% above the continuum and the absorption core is not deep enough to affect the Gaussian fits (i.e. the absorption does not go below the continuum).

\section{He I Radial Velocity Curves}
\label{sec:helium_rv}
\begin{figure}
     \centering
     \includegraphics[scale=0.3]{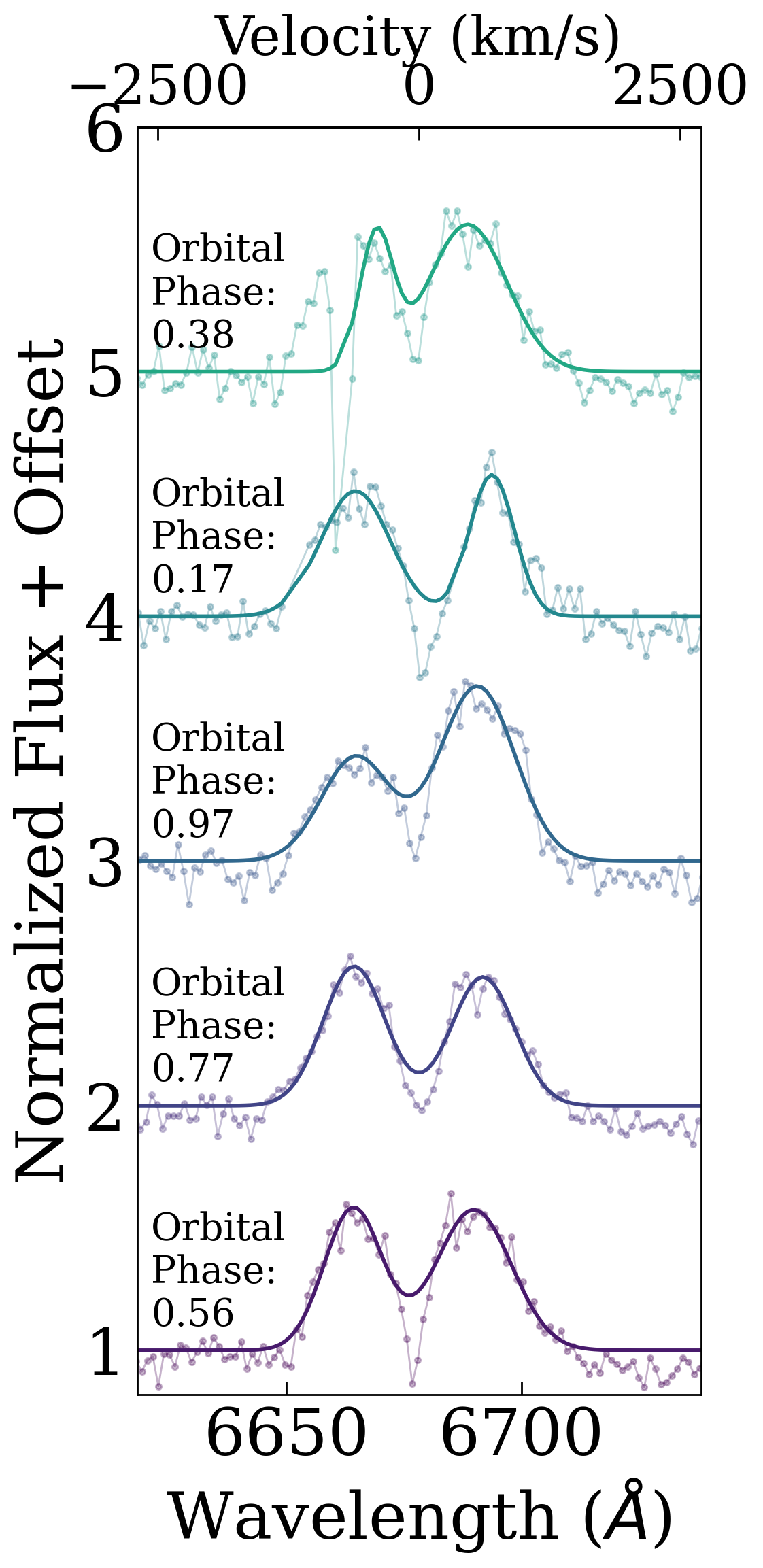}\includegraphics[scale=0.4]{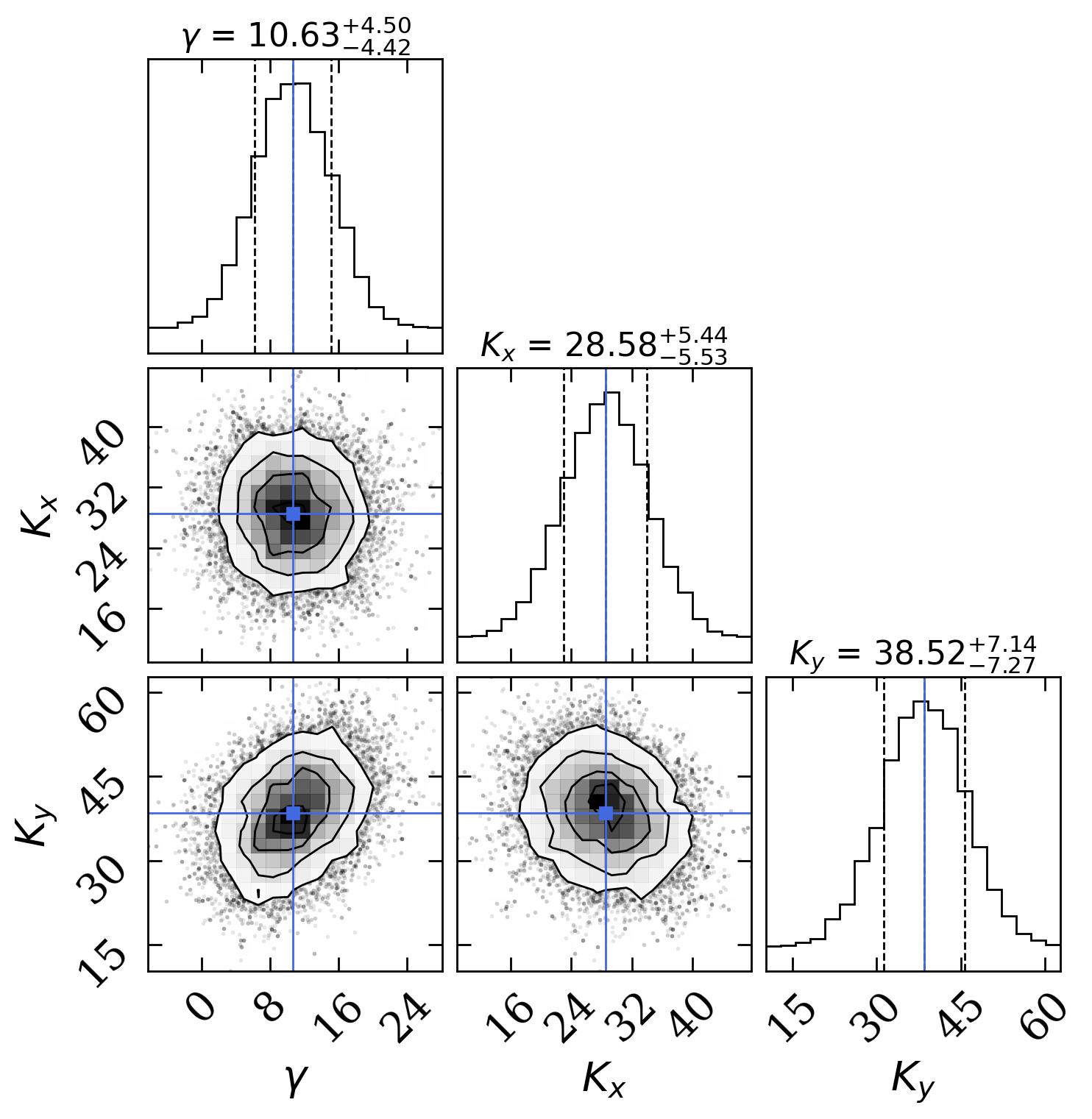}\includegraphics[scale=0.25]{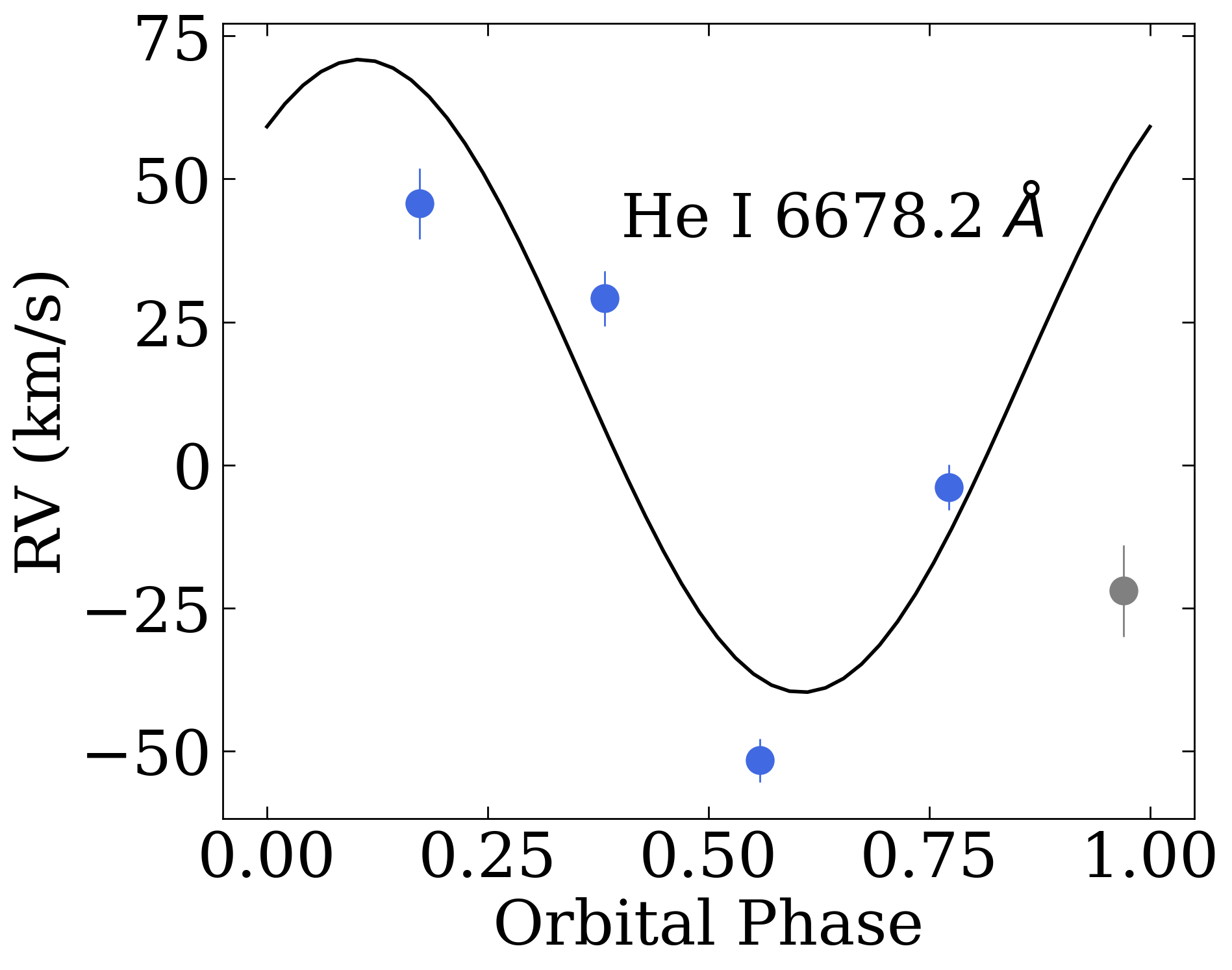}\\
     \includegraphics[scale=0.3]{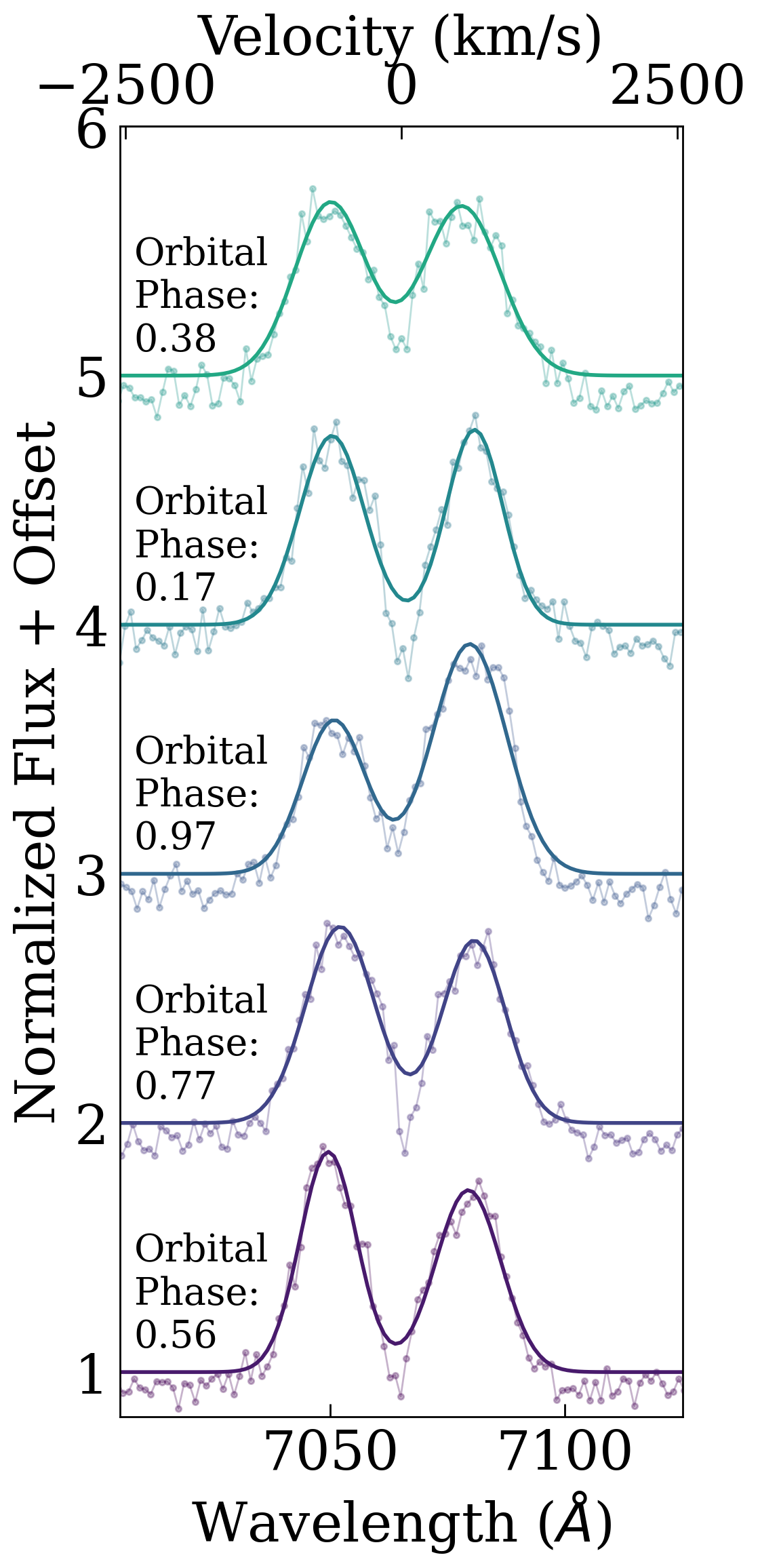}\includegraphics[scale=0.4]{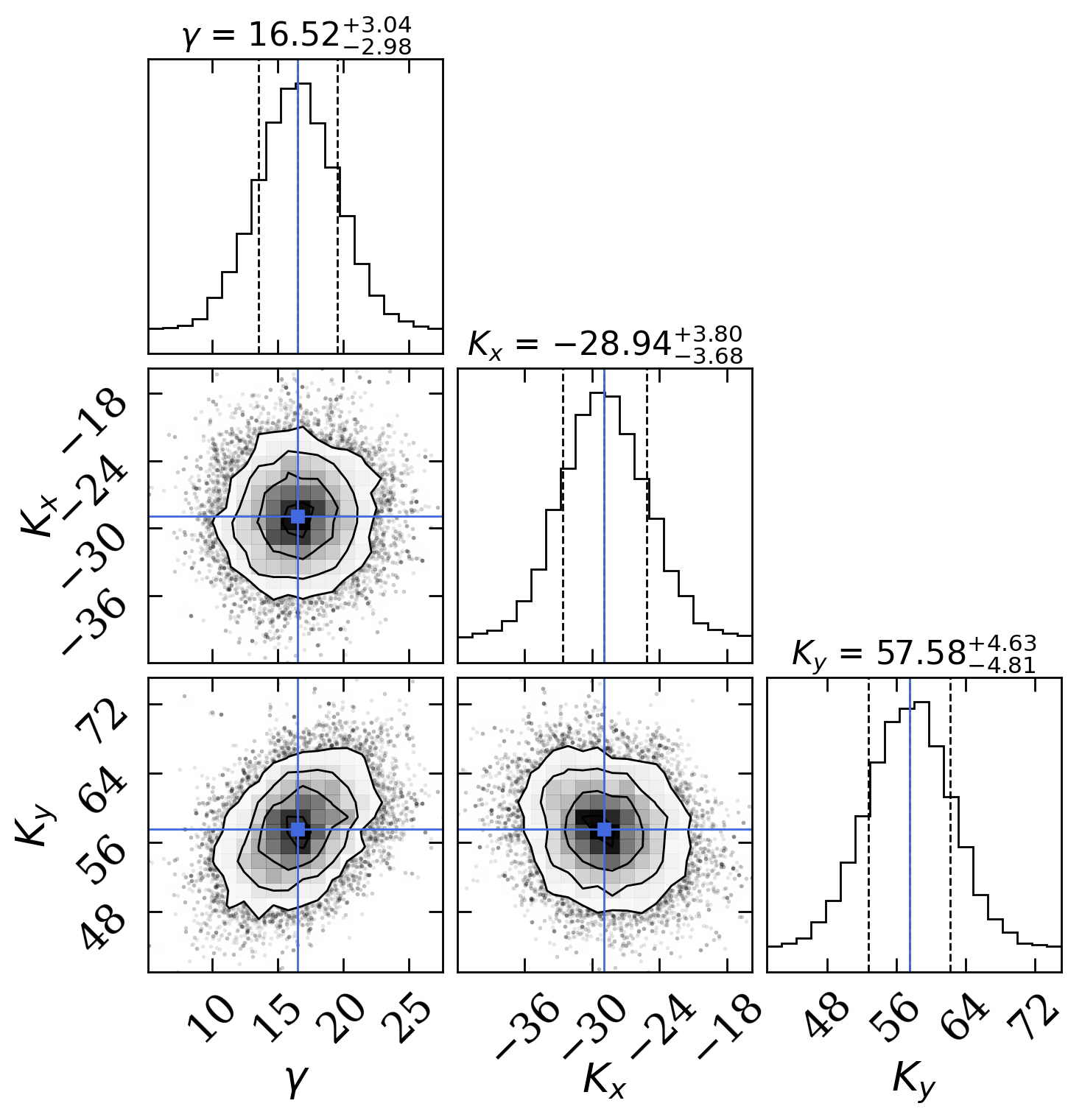}\includegraphics[scale=0.25]{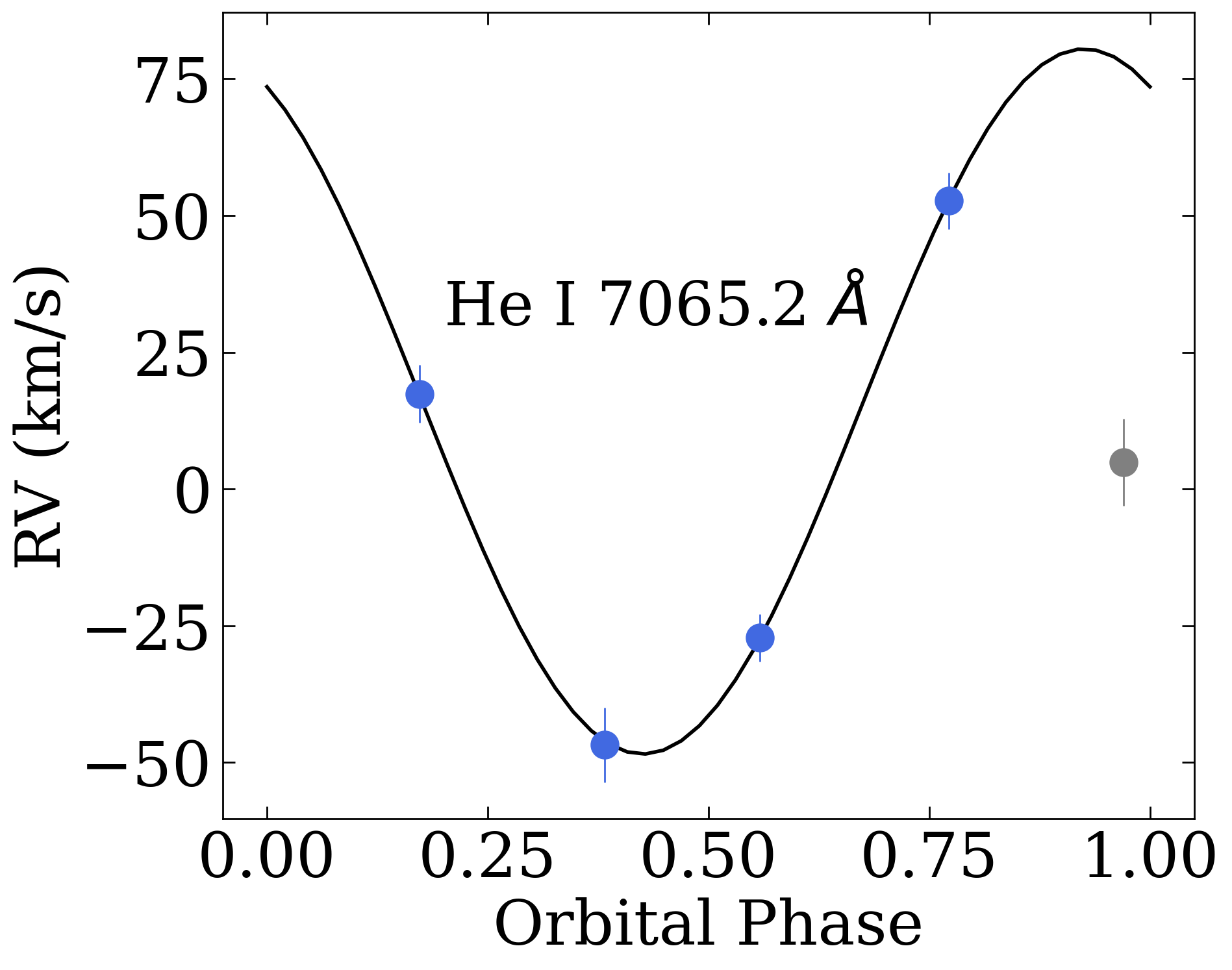}\\
     \caption{Phase-resolved spectra of He I 6678.2 \AA\; (above) and 7065.2 \AA\; (below) with resulting MCMC parameter estimates of radial velocities. The spectrum at phase 0.97  is clearly asymmetric in both lines, confirming the disk is eclipsed by the donor at the orbital phase. The MCMC parameter estimates are therefore done without the spectrum at that orbital phase, and result in a good fit that would not be possible if that point were included. The in-eclipse RV measurement is shown in grey to indicate it is not used in the MCMC analysis.}
     \label{fig:helium_i_lines_good}
 \end{figure}

We present phase-resolved spectroscopy of the He I 6678.2 and 7065.2 \AA\; lines, along with corresponding MCMC corner plots and RV curves in Figure \ref{fig:helium_i_lines_good}. Our process is the following: we independently fit two Gaussians (letting the amplitude, mean, and variance be free for each) to the two peaks of a He I emission line. Using the \texttt{curve\_fit} routine from \texttt{scipy}, we obtain the best-fit parameters and covariance matrix for each of the two fits. To compute radial velocity of a given line, we take the average of the two peaks as the central wavelength of the line at that orbital phase. 

Furthermore, we argue that spectra taken near the WD eclipse (phase $\approx0.9-1.1$) can lead to an incorrect RV analysis. From the photometry, we know that there is an eclipse, at which point both the WD and the disk are eclipsed. Since emission lines originate in the disk surrounding the WD, at this orbital phase the donor star eclipses the disk in an asymmetric manner with respect to our line of sight. In other words, the donor obscures the part of the disk where the lines are blueshifted, causing a velocity distribution that does not reflect the actual disk velocity.

Note: there is a clear feature in the He I 6678 \AA\; line at phase 0.38, with a strong absorption feature in the left peak. It is unclear if this is a cosmic ray or a real feature. The Gaussian fit to that peak does not change significantly if the affected points are omitted, and we keep the feature there for transparency.

 \section{Doppler Tomogram for Other He I Lines}
 \label{sec:dop_5876.5}
 
We present a Doppler tomogram and trailed spectra for the strongest He I line at 5875.6 \AA. Due to the strong blending with Na I, we cannot perform an adequate RV analysis to find the systemic velocity of the line, $\gamma$. In making this Doppler tomogram, we use $\gamma=17$ km/s, used for the He I 7065.2 \AA \;line. The accretion disk is seen in the Doppler tomogram along with a prominent bright spot 45 degrees ahead of the donor star (second quadrant). There is no strong evidence of another bright spot located 135--180 degrees ahead of the donor star (third quadrant of the plot). The emission seen in yellow at apparently high velocities is likely due to blending with the Na I doublet at 5890.0 and 5895.9 \AA.

We also present the Doppler tomogram and trailed spectra of the He I 6678.2 \AA\; line. The Doppler tomogram shows strong evidence there could be another bright spot located 135 degrees ahead of the donor star (third quadrant of the plot). In Figure \ref{fig:helium_i_lines_good}, we show that one of the spectra of this line could be affected by a cosmic ray, which may artificially lead to a second bright spot. Overall, the He I 6678.2 \AA\; tomogram reveals the disk to be more patchy, but further spectra are needed to verify the true structure of the disk. In short, because we do not see strong evidence for a second bright spot in \textit{all} He I emission lines, we do not conclusively determine if it exists or not.

\begin{figure}
    \centering
    \includegraphics[scale=0.2]{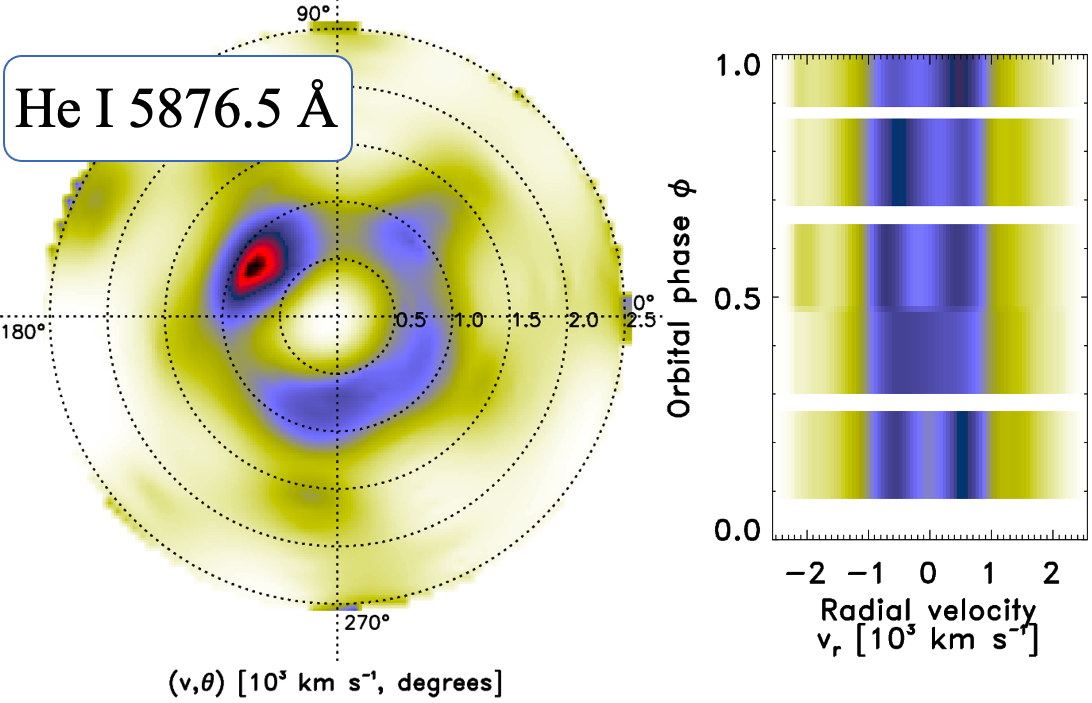}\\
    \includegraphics[scale=0.2]{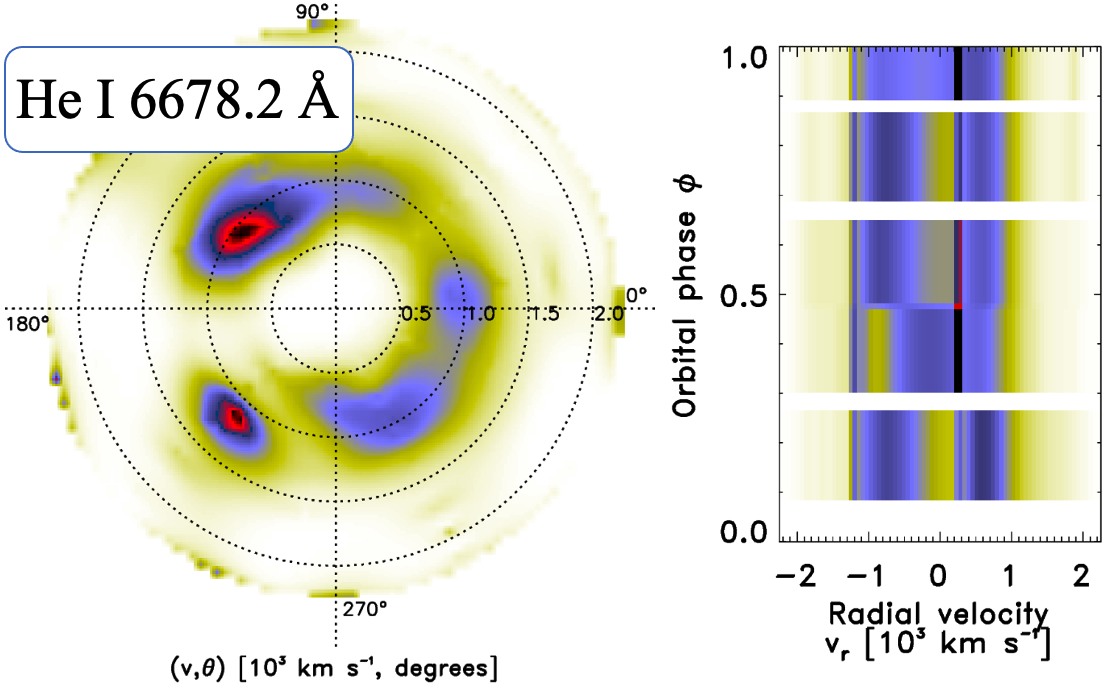}\\
    \caption{Doppler tomogram and trailed spectra for He I 5876.5  \AA\;reveals a disk with at least one prominent bright spot. The Doppler tomogram of He I 6678.2  \AA\; suggests there could be a second bright spot, but this could be an artifact due to a possible cosmic ray in one of the spectra.}
    \label{fig:doppler}
\end{figure}

\end{document}